\documentclass[a4paper,fleqn,usenatbib]{mnras}

\usepackage{newtxtext,newtxmath}

\usepackage[T1]{fontenc}
\usepackage{ae,aecompl}

\usepackage{graphicx}	
\usepackage{amsmath}	
\usepackage{amssymb}	

\title[PAHs and extinction curves in galaxy evolution]{Self-consistent modelling of
aromatic dust species and extinction curves in galaxy evolution}

\author[H. Hirashita and M. S. Murga]{
Hiroyuki Hirashita$^1$\thanks{E-mail: hirashita@asiaa.sinica.edu.tw} and
Maria S. Murga$^2$
\\
$^{1}$Institute of Astronomy and Astrophysics, Academia Sinica,
Astronomy-Mathematics Building, AS/NTU,
No.\ 1, Sec.\ 4,\\  Roosevelt Road, Taipei 10617, Taiwan \\
$^{2}$Institute of Astronomy, Russian Academy of Sciences, Pyatnitskaya str.\ 48, Moscow 119017, Russia}

\date{Accepted XXX. Received YYY; in original form ZZZ}

\pubyear{2020}

\begin{document}
\label{firstpage}
\pagerange{\pageref{firstpage}--\pageref{lastpage}}
\maketitle

\begin{abstract}
We formulate and calculate the evolution of dust in a galaxy focusing on the distinction
among various dust components -- silicate, aromatic carbon, and non-aromatic carbon.
We treat the galaxy as a one-zone object and adopt the evolution model of grain size distribution
developed in our previous work. We further include aromatization and aliphatization (inverse
reaction of aromatization). We regard small aromatic grains
in a radius range of 3--50 \AA\ as polycyclic aromatic hydrocarbons (PAHs).
We also calculate extinction curves in a consistent manner with the abundances of silicate and
aromatic and non-aromatic carbonaceous dust.
Our model nicely explains the PAH abundance as a function of metallicity in nearby galaxies.
The extinction curve become similar to the Milky Way curve at age $\sim$ 10 Gyr,
in terms of the carbon bump strength and the far-ultraviolet slope.
We also apply our model to starburst galaxies by shortening the star formation time-scale
(0.5 Gyr) and
increasing the dense-gas fraction (0.9), finding that the extinction curve
maintains bumpless shapes
(because of low aromatic fractions), which are similar to the extinction curves observed in
the Small Magellanic Cloud and high-redshift quasars.
Thus, our model successfully explains the variety in extinction curve shapes at
low and high redshifts.
\end{abstract}

\begin{keywords}
dust, extinction -- galaxies: evolution
-- galaxies: ISM -- galaxies: starburst -- molecular processes -- solid state: refractory
\end{keywords}

\section{Introduction}\label{sec:intro}

Dust grains in galaxies absorb stellar ultraviolet (UV)--optical light and
reprocess it in the infrared. This means that dust modifies or even
governs the
spectral energy distributions (SEDs) of galaxies. As explained below,
the wavelength dependence and efficiency of absorption (or extinction) and emission
depend on the dust properties. In particular, carbonaceous species
have some prominent features both in the extinction and the emission.

The infrared (IR) regime is important for tracing the physical states of the
interstellar medium (ISM) through the dust emission. In particular,
mid-infrared (MIR) spectral energy distributions (SEDs)
of galaxies usually have prominent emission features, some of which
are considered to be caused by carbonaceous species likely to be
polycyclic aromatic hydrocarbons (PAHs) \citep{Leger:1984aa,Allamandola:1985aa,Li:2012aa}.
The luminosities of these emission bands {could be a good indicator of star formation
activities} in galaxies \citep{Forster:2004aa,Peeters:2004aa}, probably because PAHs are excited by
UV radiation originating from young massive stars.
PAHs {also affect} the energy balance of the ISM {via photoprocesses}
and the ionization balance via interaction with {charged particles}
\citep[e.g.][]{Tielens:2008aa}.

Because of the above importance, the carriers of the MIR emission features have
been extensively investigated. Although PAHs are probable carriers,
there are some alternative {possible carriers}
such as hydrogenated amorphous carbons (HAC; \citealt{Duley:1993aa}),
quenched carbonaceous composite \citep{Sakata:1984aa},
and mixed aromatic/aliphatic organic nanoparticles \citep{Kwok:2011aa}.
{We adopt PAHs to represent the carriers in this paper since the scenario and
conclusion are not affected by material details}.

Small carbonaceous grains are also considered to be important in dust extinction
of UV light.
Carbonaceous dust may imprint a prominent feature also in extinction curves through
the so-called 2175 \AA\ bump. Small graphite grains are candidate carriers for
this bump \citep{Stecher:1965aa,Gilra:1971aa}.
\citet{Mathis:1994aa}
proposed that PAHs be responsible for the bump
\citep[see also][]{Weingartner:2001aa,Li:2001aa,Steglich:2010aa}.
\citet{Jones:2013aa} adopted hydrocarbon solids for the carbonaceous component.
In their model, the 2175 \AA\ bump is contributed from dehydrogenated amorphous carbon,
which is aromatic-rich. In all the above studies, the 2175 \AA\ bump is commonly attributed
to organized carbon structures like aromatic carbons and graphite.

Since the PAH abundance (or more precisely,
the luminosity of the MIR emission features) depends strongly on the metallicity
\citep[e.g.][]{Engelbracht:2005aa}, we expect that the formation and evolution of small carbonaceous
grains have a strong link to galaxy evolution.
The MIR features are deficient in low-metallicity galaxies with active star formation
\citep[e.g.][]{Hunt:2010aa}.
There are some explanations for this deficiency.
Enhanced supernova (SN) rates
could {raise the PAH destruction efficiencies} in low-metallicity environments
\citep{OHalloran:2006aa}.
It has also been suggested that strong and/or hard UV radiation field is important for PAH destruction
\citep{Madden:2000aa,Plante:2002aa,Madden:2006aa,Wu:2006aa}.
However, according to \citet{Sandstrom:2012aa}, the size distribution of PAHs are
{biased} to small sizes
in the Small Magellanic Cloud (SMC), which
may be in tension with the destruction scenarios predicting the opposite trend.

The metallicity dependence of PAH abundance may also indicate that PAH
evolution is strongly linked to the dust and metal enrichment.
This dependence is also seen at
high redshift ($z\sim 2$; \citealt{Shivaei:2017aa}); thus, it is important to
{investigate the physical mechanism that produces
the strong PAH--metallicity relation} in the context of galaxy evolution.
\citet{Galliano:2008aa} suggested that
young galaxies do not have sufficient time for low-mass stars to
evolve into asymptotic giant branch (AGB) stars, which produce
carbonaceous dust and PAHs
\citep[see also][]{Bekki:2013aa}.
{However, the deficiency of PAHs in low-metallicity galaxies may not be simply
attributed to the age effect
since there is no one-to-one correspondence between age and metallicity for various types of
galaxies}
\citep[e.g.][]{Kunth:2000aa}.

\citet{Seok:2014aa} provided another explanation for the metallicity dependence of PAH
abundance. Their model assumed that the production of small grains by shattering
is the source of PAHs. They simply regarded small carbonaceous grains as PAHs.
Since the efficiency of shattering depends strongly on the dust (and metal) abundance,
their model naturally explains the relation between PAH abundance and metallicity.
\citet[][hereafter R19]{Rau:2019aa} directly treated the evolution of grain size distribution
and the aromatization reaction by UV radiation, regarding small aromatic grains as PAHs.
They post-processed hydrodynamic simulation data in \citet[][hereafter HA19]{Hirashita:2019aa}
(originally from \citealt{Aoyama:2017aa}) and
consistently treated the dependence of dust processing on the physical condition of the ISM.
As a consequence, they not only successfully explained the relation between PAH abundance
and metallicity but also showed that the nonlinear metallicity dependence of
PAH abundance can be reproduced by the rapid increase of small grains as a result of the interplay
between shattering
and accretion.

The success of the above PAH evolution models implies that they
can also be used to predict in general small carbonaceous grains, which contribute
to the prominent 2175 \AA\ feature in the Milky Way extinction curve.
Moreover, since it is known that the extinction curves are different among galaxies
\citep[e.g.][]{Gordon:2003aa},
the evolution model of extinction curve would be useful to understand what causes
such a variation.
Indeed, a treatment of aromatic and non-aromatic components
(or carbons with ordered and disordered structures) is useful in modelling
extinction curves since, as mentioned above, the 2175 \AA\ bump is related to
carbonaceous materials with
ordered structures such as graphite and PAHs.
Therefore, it would be interesting to predict the evolution of extinction curve
by extending the above PAH evolution models.
We also include silicate as well as carbonaceous components for this purpose.
IR SEDs could also be predicted, but to do this, further detailed modelling of
stellar radiation as a heating source of dust \citep[e.g.][]{Draine:2001aa}
and/or radiation transfer calculations \citep[e.g.][]{Bianchi:2000ab,Baes:2011aa} would be needed.
Therefore, in this paper, we focus on
extinction curves and leave IR SEDs for future work.

As shown by R19, aromatization occurs on a shorter time-scale than shattering and
accretion, which justifies \citet{Seok:2014aa}'s assumption of small carbonaceous grains being
PAHs. As a result of aromatization by photoprocessing (as also experimentally shown by
\citealt{Duley:2015aa}),
carbonaceous grains which have predominantly disordered atomic structures with aliphatic bonds,
are processed by dissociation of C--H bonds. As a consequence, those grains obtain
aromatic bonds with ordered atomic structures.
In fact, the inverse reaction -- aliphatization -- also occurs in the dense ISM as a result of
the accretion of hydrogen and other atoms. Moreover, aromatization could also take place in SN shocks
\citep{Murga:2019aa}. Therefore, in this paper, we also aim at comprehensive
modelling of aromatization and aliphatization.

In summary, we model the evolution of grain size distribution, laying particular
emphasis on the aromatic component in this paper. The evolution of aromatic component is tested against
the observed PAH abundances. Moreover, by including all the dust components
(silicate, aromatic carbon, and non-aromatic carbon), we also predict the evolution of
extinction curves.
As a consequence, we complete a model that provides a way to predict the observed dust
properties. In particular, this enables us to obtain
(i) an understanding of various shapes of extinction curves observed in a variety of galaxies
and (ii) a self-consistent understanding between dust (especially grain size distribution) and PAH evolution.

This paper is organized as follows.
In Section~\ref{sec:model}, we describe the dust evolution model, which treats
the elemental compositions of relevant chemical species, the grain size distribution, and
the aromatic fraction.
In Section \ref{sec:result}, we show the results.
In Section \ref{sec:discussion}, we provide some extended discussions of the results.
In Section \ref{sec:conclusion}, we give the conclusion of this paper.

\section{Model}\label{sec:model}

To concentrate on the dust evolution, a galaxy is simply treated as a
one-zone object. Most of the components in the dust evolution model are
based on our previous formulations.
The evolution of grain size distribution is the backbone of the model.
We newly distinguish between silicate and carbonaceous dust according to the
production of silicon (Si) and carbon (C). For this purpose, we formulate the stellar
metal and dust production.
We further separate the carbonaceous dust into aromatic and non-aromatic components.
Small aromatic grains represent PAHs
in our model. For a representative observational quantity, we calculate the
extinction curve.

\subsection{Galaxy chemical evolution model}\label{subsec:chemical}

We model the galaxy evolution, especially, the chemical enrichment
as a result of stellar birth and death. The galaxy is treated as a one-zone object;
that is, we do not model its spatially resolved information. For simplicity, we assume
the galaxy to be a closed box; that is, we presume that the galaxy starts
with a gas mass with zero metallicity and converts the gas to stars with
the total baryonic (gas + stars) mass conserved.

We denote the masses of gas, stars, metals, and dust in the galaxy as
$M_\mathrm{gas}$, $M_\star$, $M_Z$, and $M_\mathrm{dust}$, respectively.
The above closed-box assumption leads to $M_\mathrm{g,0}=M_\mathrm{gas}+M_\star$,
where $M_\mathrm{g,0}$ is the total baryonic mass.
We adopt $M_\mathrm{g}=M_\mathrm{g,0}$ at time $t=0$ for the initial condition.
The gas mass and the metal mass evolve as the star formation proceeds as
\begin{align}
\frac{\mathrm{d}M_\mathrm{gas}}{\mathrm{d}t} &= -\psi (t)+R(t),\label{eq:Mgas}\\
\frac{\mathrm{d}M_{Z}}{\mathrm{d}t} &= -Z\psi (t)+Y_Z(t),\label{eq:metal}
\end{align}
where $\psi (t)$ is the star formation rate, $R(t)$ is the gas mass return rate
from the stars at their death, $Z\equiv M_Z/M_\mathrm{gas}$ is the
metallicity, and $Y_Z(t)$ is the mass ejection rate of metals from stars.
We also express the evolution of dust mass as (note that we do not use this equation directly in our
calculation)
\begin{align}
\frac{\mathrm{d}M_\mathrm{dust}}{\mathrm{d}t} &= -\mathcal{D}\psi (t)+Y_\mathrm{dust}(t)
+\dot{M}_\mathrm{dust,ISM},\label{eq:Mdust}
\end{align}
where $\mathcal{D}\equiv M_\mathrm{dust}/M_\mathrm{gas}$ is the dust-to-gas ratio,
$Y_\mathrm{dust}(t)$ is the ejection rate of dust from stars, and
$\dot{M}_\mathrm{dust,ISM}$ is the changing rate of dust mass by
interstellar processing (dust growth by accretion and dust destruction in SN shocks).
We derive the following equation for $\mathcal{D}$ by 
combining equations (\ref{eq:Mgas}) and (\ref{eq:Mdust}):
\begin{align}
\frac{\mathrm{d}\mathcal{D}}{\mathrm{d}t}=\frac{Y_\mathrm{dust}-\mathcal{D}R}{M_\mathrm{gas}}
+\frac{\dot{M}_\mathrm{dust,ISM}}{M_\mathrm{gas}}.
\end{align}
We define the contribution from stellar dust production to $\mathcal{D}$ as
$\mathcal{D}_\star$, which is calculated as
\begin{align}
\frac{\mathrm{d}\mathcal{D}_\star}{\mathrm{d}t}=
\frac{Y_\mathrm{dust}-\mathcal{D}_\star R}{M_\mathrm{gas}}.\label{eq:dg_star}
\end{align}
Note that the second term on the right-hand side is due to the dilution of dust-to-gas ratio
by gas ejection.
We use $\mathrm{d}\mathcal{D}_\star /\mathrm{d}t$ to estimate the
contribution from the stellar dust production to the increase of dust abundance
(see Section~\ref{subsec:size}).

Similarly, the metallicity $Z\equiv M_Z/M_\mathrm{gas}$ is governed by the following
equation, which can be derived by combining equations (\ref{eq:Mgas}) and (\ref{eq:metal}):
\begin{align}
\frac{\mathrm{d}Z}{\mathrm{d}t}=\frac{Y_Z-ZR}{M_\mathrm{gas}}.\label{eq:metallicity}
\end{align}

We assume the following simple functional form for the star formation rate:
\begin{align}
\psi (t)=M_\mathrm{g,0}\exp (-t/\tau_\mathrm{SF}),
\end{align}
where $\tau_\mathrm{SF}$ is the star formation time-scale given as a free parameter.
We assume the exponential decline for a smooth shutdown of star formation on
a time-scale of $\tau_\mathrm{SF}$.
The other necessary functions are evaluated as
\begin{align}
R(t) &= \int_{m_t}^{m_\mathrm{u}}[\tilde{m}-w(\tilde{m},\, Z(t-\tau_{\tilde{m}} ))]\,\phi (\tilde{m})\,
\psi(t-\tau_{\tilde{m}})\,\mathrm{d}\tilde{m},\\
Y_Z(t) &= \int_{m_t}^{m_\mathrm{u}}m_Z(\tilde{m},\, Z(t-\tau_{\tilde{m}} ))\,\phi (\tilde{m})\,
\psi(t-\tau_{\tilde{m}})\,
\mathrm{d}\tilde{m},\label{eq:metal_yield}\\
Y_\mathrm{dust}(t) &= \int_{m_t}^{m_\mathrm{u}}m_\mathrm{d}(\tilde{m},\, Z(t-\tau_{\tilde{m}} ))\,\phi
(\tilde{m})\,\psi(t-\tau_{\tilde{m}})\,\mathrm{d}\tilde{m},
\end{align}
where $m_t$ is the turn-off mass, $\tau_{\tilde{m}}$ is the lifetime of a star with mass $\tilde{m}$
(note that $\tau_{{m}_t}=t$; the stellar mass is defined at the zero age main sequence),
$m_\mathrm{u}$ is the upper stellar mass limit
(we adopt 100~M$_{\sun}$), $\phi (\tilde{m})$ is the initial mass function (IMF),
$w(\tilde{m},\, Z)$ is the remnant mass as a function of stellar mass and
stellar metallicity, $m_Z(\tilde{m},\, Z)$ and $m_\mathrm{d}(\tilde{m},\, Z)$ are
the mass of metals and dust, respectively, produced by a star with
mass $\tilde{m}$ and metallicity $Z$
(note that $w$, $m_Z$, and $m_\mathrm{d}$ are evaluated with the ISM metallicity at the time when
the star is formed).
We adopt the stellar lifetime from \citet{Raiteri:1996aa}.
We consider core-collapse SNe and AGB stars, which are known to be the main
contributors for stellar dust production \citep{Dwek:1998aa}. We assume that the progenitor masses of
AGB stars and SNe are $\tilde{m}<8$~M$_{\sun}$ and $8<\tilde{m}<40$~M$_{\sun}$,
respectively. We simply assume that stars with $\tilde{m}>40$ M$_{\sun}$ collapse into
black holes without ejecting any gas, metals or dust \citep{Heger:2003aa}.
The metal masses produced by AGB stars and SNe are taken from \citet{Karakas:2010aa}
and \citet{Kobayashi:2006aa}, respectively.
The remnant masses are also adopted from the same papers.
We adopt the Chabrier IMF \citep{Chabrier:2003aa} for $\phi (\tilde{m})$
with a stellar mass range of 0.1--100 M$_{\sun}$.

For the dust evolution calculated later, we also need the abundances of silicon
(Si) and carbon (C). The mass abundances of Si and C are denoted as
$Z_\mathrm{Si}$ and $Z_\mathrm{C}$, respectively. These values are used to derive the
fractions of silicate and carbonaceous dust (Section \ref{subsec:separation}).
Both of them are calculated by using equation (\ref{eq:metallicity}),
but by replacing the stellar yield data of the total metal mass with those of Si and C
in equation (\ref{eq:metal_yield}). The yield data of Si and C are available in the same
references as those of the total metal mass.

We also calculate the SN rate, $\gamma (t)$, by
\begin{align}
\gamma (t)=\int_{8~\mathrm{M}_{\sun}}^{m_\mathrm{u}}\phi
(\tilde{m})\,\psi(t-\tau_{\tilde{m}})\,\mathrm{d}\tilde{m}.\label{eq:SNR}
\end{align}
This is necessary to estimate the dust destruction rate in Section~\ref{subsubsec:destruction}
and the aromatization rate in SN shocks in Section \ref{subsec:separation}.

\subsection{Evolution of grain size distribution}\label{subsec:size}

We adopt the model for the evolution of grain size distribution from HA19.
For the dust evolution processes,
we consider stellar dust production, dust destruction by SN shocks in
the ISM, dust growth by accretion and coagulation in the dense ISM, and
dust disruption by shattering in the diffuse ISM. We only describe the outline,
and refer the interested reader to HA19 for further details.

We aim at separating the dust species; however, because of grain--grain
collisions among different dust species,
the evolution of grain size distributions for
multiple species is complicated. For example, grain growth could form
compound species or result in a core-mantle structure. Moreover,
whether or not such compound species survive
robustly is not clear because the binding force at the interface of
different species could be weak.
Thus, for simplicity, we neglect such an `inter-species' complexity:
we first calculate the total grain size distribution for all the dust species,
and later divide the calculated grain size distribution into the
individual dust species. This also means that we neglect the different efficiencies of
various dust processing mechanisms between silicate and carbonaceous dust.
We use graphite properties for the calculation of the total grain
size distribution. We also show the calculation with silicate material properties
later to test the robustness (Section \ref{subsec:robustness}).

We assume grains to be spherical and compact,
so that $m=(4\upi /3)a^3s$, where $m$ is the grain mass, $a$ is the grain radius and $s$
is the material density of dust.
We adopt $s=2.24$ g cm$^{-3}$
(for graphite; $s=3.5$ g cm$^{-3}$ for silicate)
\citep{Weingartner:2001aa}.
The grain size distribution at time $t$ is expressed by the grain mass distribution
$\rho_\mathrm{d}(m,\, t)$, which is defined such that
$\rho_\mathrm{d}(m,\, t)\,\mathrm{d}m$ is the mass density
of dust grains whose mass is between $m$ and $m+\mathrm{d}m$.
The grain mass distribution is related
to the grain size distribution,
$n(a,\, t)$, as
\begin{align}
\rho_\mathrm{d}(m,\, t)\,\mathrm{d}m=\frac{4}{3}\upi a^3sn(a,\, t)\,\mathrm{d}a.\label{eq:rho_vs_n}
\end{align}

We also define the grain mass distribution per gas density
as $\tilde{\rho}(m,\, t)\equiv\rho(m,\, t)/\rho_\mathrm{gas}$, where
$\rho_\mathrm{gas}$ is the gas density given by the number density of hydrogen nuclei,
$n_\mathrm{H}$, as
$\rho_\mathrm{gas}=\mu m_\mathrm{H}n_\mathrm{H}$
($\mu =1.4$ is the gas mass per hydrogen, and $m_\mathrm{H}$ is the mass
of hydrogen atom).
The dust-to-gas ratio is estimated by
\begin{align}
\mathcal{D}(t)=\int_0^\infty\tilde{\rho}_\mathrm{d}(m,\, t)\,\mathrm{d}m.
\end{align}
We explain how to set $n_\mathrm{H}$ in what follows.

We consider that the ISM is composed of the diffuse (warm) and dense (cold) components,
which have
$(n_\mathrm{H}/\mathrm{cm}^{-3},\, T_\mathrm{gas}/\mathrm{K})=(0.3,\, 10^4)$
and $(300,\, 25)$, respectively (\citealt{Nozawa:2015aa}; HA19),
where $T_\mathrm{gas}$ is the gas temperature.
The mass fraction of the dense component is denoted as
$\eta_\mathrm{dense}$; the diffuse component occupies a mass fraction of $1-\eta_\mathrm{dense}$.
The information on the ISM phase is used in the following way.
We calculate the change of grain mass distribution
$\Delta\tilde{\rho}_\mathrm{d}(m,\, t)=[{\upartial\tilde{\rho}_\mathrm{d}(m,\, t)}/{\upartial t}]_i\,f_i\Delta t$,
where $i$ indicates each process, that is,
$[{\upartial\tilde{\rho}_\mathrm{d}(m,\, t)}/{\upartial t}]_i$ is the contribution from
process $i$ to the change of the grain mass distribution,
and $f_i$ is the fraction of the gas phase that host the process.
Since our model cannot treat the spatial distribution of the gas, the fraction of the gas phase
is included by weight $f_i$.
We consider stellar dust production ($i=\mbox{star}$), SN destruction by sputtering
($i=\mbox{sput}$),
grain disruption by shattering ($i=\mbox{shat}$), dust growth by accretion ($i=\mbox{acc}$),
and grain growth by coagulation ($i=\mbox{coag}$).
Stellar dust production and SN destruction are assumed to occur in both
ISM phases, so that $f_i=1$. Coagulation and accretion take place only in the
dense phase, so that $f_i=\eta_\mathrm{dense}$.
Shattering happens only in the diffuse phase, so that
$f_i=1-\eta_\mathrm{dense}$. 

In computing the grain size distribution, we discretize the entire grain radius range
($a=3\times 10^{-4}$--10 $\micron$) into
$N_\mathrm{g}=128$ grid points.
We set $\tilde{\rho}_\mathrm{d}(m,\, t)=0$ at the maximum and minimum grain radii for
the boundary conditions.

\subsubsection{Stellar dust production}\label{subsec:stellar}

In our previous model (HA19), we fixed the dust condensation efficiency in stellar ejecta.
Since we calculate the dust abundance using the dust condensation calculations
referred to in Section \ref{subsec:chemical},
it is not any more necessary to assume the dust condensation efficiency.
We utilize the stellar dust production calculated
by equation (\ref{eq:dg_star}), and
write the change of the grain
size distribution by stellar dust production as
\begin{align}
\left[\frac{\upartial\tilde{\rho}_\mathrm{d}(m,\, t)}{\upartial t}\right]_\mathrm{star}=
\frac{\mathrm{d}\mathcal{D}_\star}{\mathrm{d}t}\, m\tilde{\varphi} (m),\label{eq:stellar}
\end{align}
where $m\tilde{\varphi} (m)$ is the mass distribution function of
the dust grains produced by stars, and it is normalized so that the integration
for the whole grain mass range is unity.
This grain size distribution is related to the above mass distribution as
$\varphi (a)\,\mathrm{d}a\equiv\tilde{\varphi}(m)\,\mathrm{d}m$.
For the grain size distribution of dust produced by stars, we adopt a lognormal
function with a central radius of 0.1 $\micron$ and a standard deviation of 0.47
\citep{Asano:2013aa}.

\subsubsection{Dust destruction and growth}\label{subsubsec:destruction}

{For dust destruction by sputtering and dust growth by accretion,
the time evolution of grain mass distribution is
solved in a manner consistent with the grain-size-dependent destruction and growth
rates, respectively, using an `advection' equation in the grain-radius space
(HA19).}
The destruction time-scale $\tau_\mathrm{dest}(m)$ is estimated as
\citep[e.g.][]{McKee:1989aa}:
\begin{align}
\tau_\mathrm{dest}(m)=
\frac{M_\mathrm{gas}}{\epsilon_\mathrm{dest}(m)M_\mathrm{s}\gamma},
\label{eq:tau_dest}
\end{align}
where 
$M_\mathrm{s}=6800$~M$_{\sun}$ is the gas mass swept by a single SN
blast, $\gamma$ is the SN rate calculated by equation (\ref{eq:SNR}), and
$\epsilon_\mathrm{dest}(m)$ is the dust destruction efficiency as a function of the grain mass.
We adopt an empirical expression for the destruction efficiency
(described as a function of $a$ instead of $m$) as
$\epsilon_\mathrm{dest}(a)=1-\exp [-0.1({a}/{0.1~\micron})^{-1}]$
(HA19; \citealt{Aoyama:2020aa}).
For accretion, the growth time-scale is
estimated as (we fixed the sticking efficiency $S=0.3$ in HA19)
\begin{align}
\tau_\mathrm{acc}(m) &= \frac{1}{3} \tau_\mathrm{0,acc}\left(\frac{a}{0.1~\micron}
\right)\left(\frac{Z}{\mathrm{Z}_{\sun}}\right)^{-1}
\left(\frac{n_\mathrm{H}}{10^3~\mathrm{cm}^{-3}}
\right)^{-1}\left(\frac{T_\mathrm{gas}}{10~\mathrm{K}}
\right)^{-1/2},
\label{eq:tau_acc}
\end{align}
where $\tau_\mathrm{0,acc}$ is a constant.
We adopt $\tau_\mathrm{0,acc}=0.993\times 10^8$ yr
appropriate for graphite ($\tau_\mathrm{0,acc}=1.61\times 10^8$ yr for
silicate;
\citealt{Hirashita:2012aa}).

\subsubsection{Shattering and coagulation}\label{subsubsec:shattering}

{The time evolution of grain size distribution by shattering and coagulation
is expressed by a Smoluchowski equation (HA19).
The grain--grain collision rates for various combinations of grain radii are
evaluated based on the geometric cross-section and the grain velocities.
The grain velocities are evaluated from a simple analytical model of turbulence
\citep{Ormel:2009aa} but the normalization is adjusted to effectively
realize the high grain velocities suggested by \citet{Yan:2004aa} for shattering.
The direction of the relative velocity in each collision is chosen randomly.
For shattering, the fragment mass distribution is determined following
\citet{Kobayashi:2010aa} using the tensile strength appropriate for compact
grains [in their notation, we adopt $Q_\mathrm{D}^\star =8.9\times 10^{9}$ cm$^2$ s$^{-2}$
(valid for graphite; for silicate,
$Q_\mathrm{D}^\star =4.3\times 10^{10}$ cm$^2$ s$^{-2}$)].
The maximum and minimum masses of the fragments
are assumed to be
$m_\mathrm{f,max}=0.02m_\mathrm{ej}$ and
$m_\mathrm{f,min}=10^{-6}m_\mathrm{f,max}$, respectively \citep{Guillet:2011aa}.
We adopt the following mass distribution function of grain $m_1$ in the collision with
a grain with mass $m_2$
as\footnote{{In HA19, the expression (their equation 25) is valid
only for $m_\mathrm{f,min}\leq m\leq m_\mathrm{f,max}$,
so we write a mathematically precise form applicable to any grain mass
(note that we used this precise form also in HA19).}}
\begin{align}
\mu_\mathrm{shat}(m,\, m_1,\, m_2) &=
\frac{(4-\alpha_\mathrm{f})m_\mathrm{ej}m^{(-\alpha_\mathrm{f}+1)/3}}{3\left[
m_\mathrm{f,max}^\frac{4-\alpha_\mathrm{f}}{3}-
m_\mathrm{f,min}^\frac{4-\alpha_\mathrm{f}}{3}\right]}\,
\Phi (m;\, m_\mathrm{f,min},\, m_\mathrm{f,max})\nonumber\\
&+ (m_1-m_\mathrm{ej})\delta (m-m_1+m_\mathrm{ej}),\label{eq:frag}
\end{align}
where $m_\mathrm{ej}$ is the total fragment mass ejected from $m_1$ (which depends on the
colliding grain mass $m_2$ and the relative velocity),
$\Phi (m;\, m_\mathrm{f,min},\, m_\mathrm{f,max})=1$ if
$m_\mathrm{f,min}\leq m\leq m_\mathrm{f,max}$, and 0 otherwise,
$\delta (\cdot )$ is Dirac's delta function, and
$\alpha_\mathrm{f}=3.3$ \citep{Jones:1996aa}.
Grains which become smaller than the minimum grain size
($a=3\times 10^{-4}~\micron$)
are removed.
For coagulation, we assume the sticking efficiency to be unity.}

\subsection{Separation into various species}\label{subsec:separation}

Now, we decompose
the grain size distribution into the relevant dust species.
For simplicity, we assume that
silicate and carbonaceous species have the same grain size
distribution; that is, the grain mass distributions of silicate ($\rho_\mathrm{sil}$) and
carbonaceous dust ($\rho_\mathrm{car}$) are
described by
\begin{align}
\rho_\mathrm{sil}(m,\, t) &= f_\mathrm{sil}(t)\rho_\mathrm{d}(m,\, t),\\
\rho_\mathrm{car}(m,\, t) &= [1-f_\mathrm{sil}(t)]\rho_\mathrm{d}(m,\, t),
\end{align}
where $f_\mathrm{sil}(t)$ is the mass ratio of silicate to the total dust mass,
which is independent of $m$ by the above assumption.

The silicate fraction $f_\mathrm{sil}$ is calculated by
\begin{align}
f_\mathrm{sil}(t)=\frac{6Z_\mathrm{Si}(t)}{6Z_\mathrm{Si}(t)+Z_\mathrm{C}(t)},
\label{eq:sil}
\end{align}
where the abundances of Si and C are calculated in Section~\ref{subsec:chemical}.
The factor 6 comes from the mass fraction of Si in silicate \citep{Hirashita:2011aa}.

Now we separate the carbonaceous component into aromatic and non-aromatic populations.
We denote the grain mass distribution of the aromatic species as
$\rho_\mathrm{ar}(m,\, t)$. We also introduce the aromatic fraction, which is defined as
$f_\mathrm{ar}(m,\, t)\equiv\rho_\mathrm{ar}(m,\, t)/\rho_\mathrm{car}(m,\, t)$.
We assume (i) that there are aliphatic-dominated and aromatic-dominated species,
and (ii) that aromatization converts the former to the latter ones.
For aromatization, we consider photoprocessing and processing in SN shocks.
We also include the inverse reaction of aromatization -- aliphatization.
We take aliphatization by hydrogenation and by accretion of carbon into account.
We explain these processes, of which the rates are estimated, in what follows.

Owing to photo-processing, grains lose mainly their hydrogen atoms.
Herewith, atomic structures change to aromatic bonds.
We trace the aromatization by the change of
the band gap energy, $E_{\rm g}$, which is related
to the number fraction of hydrogen atoms through $E_{\rm g}=4.3 X_{\rm H}$ eV
\citep{Tamor:1990aa}.
We assume the maximum and minimum values of $E_{\rm g}$ to be
2.67~eV (fully hydrogenated case corresponding to $X_\mathrm{H}=0.6$;
\citealt{Jones:2013aa}) and 0.1~eV (corresponding to
$X_{\rm H}=0.02$), respectively.
We define the aromatization time as the time
necessary for dehydrogenation from the maximum value of $X_{\rm H}$ to the minimum one.

For aromatization by photo-processing, we described the calculation method in R19.
We adopt the stellar radiation field from \citet{Mathis:1983aa} and scale it with a constant parameter
$U$ following \citet{Draine:2007aa} ($U=1$ corresponds to the Milky Way radiation field
in the solar neighbourhood).
Finally, we obtain the following fitting formula for the aromatization time by photo-processing:
\begin{align}
\frac{\tau_\mathrm{ar}^\mathrm{UV}}{\mathrm{yr}}=U^{-1}\left[3\left(\frac{a}{\micron}\right)^{-2} + 6.6\times 10^{7}
\left(\frac{a}{\micron}\right)\right] .
\end{align}
Ideally, we could solve the stellar spectrum in a consistent manner with the star formation history.
However, the interstellar radiation field is not solely determined by the stellar spectra:
the spatial distributions (geometries) of dust and stars affect the radiation incident on the dust
through radiation transfer effects. As shown later, the aromatization time-scale
is much shorter than the other relevant processes for most of the grain radius range.
This guarantees that the aromatic fraction converges to an equilibrium value
quickly and that the results are insensitive to the aromatization time-scale (see below).
Star-forming galaxies usually have $U\gtrsim 1$ \citep[e.g.][]{Draine:2007aa}, so that
we adopt $U=1$ in evaluating the aromatization time-scale
(to show that the equilibrium is achieved even in this conservative estimate).

UV radiation, especially hard photons, could destroy small PAHs.
We estimated the destruction time-scale
based on \citet{Murga:2019aa} and found that a hard UV SED appropriate for
low metallicity galaxies could destroy PAHs with $a\sim 3$ \AA\ (5 \AA) if $U\gtrsim 1$
(10) within the cosmic age. However, the supply of small grains by shattering occurs in a
much shorter time ($\sim 10^8$--$10^9$ yr), within which only the smallest PAHs ($a\sim 3$ \AA)
could be destroyed by photo-processing.
Therefore, we argue that PAH destruction by hard UV is negligible as far as the
general interstellar radiation field is concerned. PAHs could be destroyed locally in
regions near to massive stars, but could also be
continuously formed in the diffuse ISM by shattering. Such spatially dependent PAH destruction and
formation are hard to treat in our one-zone model. Therefore, we neglect photo-destruction
in this paper, but leave the imprint of this effect for future spatially resolved modelling.

We also consider aromatization by SNe. According to \citet{Murga:2019aa},
carbonaceous grains with $a\lesssim 0.1~\micron$ are fully aromatized in a single SN shock.
Therefore, we simply adopt the following form for the aromatization efficiency in a single
SN ($\epsilon_\mathrm{ar,SN}$):
\begin{align}
\epsilon_\mathrm{ar,SN}(a)=1-\exp\left[-\left(\frac{a}{0.1~\micron}\right)^{-1}\right] ,
\end{align}
With this efficiency, the time-scale of aromatization by SNe in the galaxy is
estimated as
\begin{align}
\tau_\mathrm{ar}^\mathrm{SN}=
\frac{M_\mathrm{gas}}{\epsilon_\mathrm{ar,SN}(m)M_\mathrm{s}\gamma},
\end{align}
where $\epsilon_\mathrm{ar,SN}$ is expressed as a function of grain mass ($m$) instead of
grain radius (note a similar form to the dust destruction by SNe in
equation \ref{eq:tau_dest}). Thus,
aromatization by SNe occurs on the SN-sweeping time-scale in the entire ISM, which is
on the order of $\sim 10^7$--$10^8$ yr. Since this is much longer than the aromatization time for
UV processing, aromatization in SN shocks generally
has a negligible impact on the aromatic fraction.

For aliphatization, we consider two processes caused by the accretion of hydrogen
and that of metals. The latter process is considered because,
when dust grains accrete carbon in the cold ISM, the accreted material
cannot get enough energy to form a regular atomic structure. Such a material with irregular
structures is categorized as a non-aromatic species in our model.
We calculated the time-scale of aliphatization by the accretion of hydrogen
($\tau_\mathrm{al}^\mathrm{H}$) based on
\citet{Murga:2019aa}, and obtained the following
fitting formula:
\begin{align}
\frac{\tau_\mathrm{al}^\mathrm{H}}{\mathrm{yr}} = 1.6\times 10^{5}\left({\frac{a}{\micron}}
\right) .
\end{align}
The time-scale of aliphatization by the accretion of metals ($\tau_\mathrm{al}^\mathrm{acc}$)
is estimated as
\begin{align}
\tau_\mathrm{al}^\mathrm{acc} =\frac{\tau_\mathrm{acc}}{\xi (t)},
\end{align}
where $\tau_\mathrm{acc}$ is the accretion time-scale evaluated in equation (\ref{eq:tau_acc}),
and $\xi$ is the fraction of metals in the gas phase (Section \ref{subsubsec:destruction}).
The resulting aromatization and aliphatization time-scales (denoted as $\tau_\mathrm{ar}$ and
$\tau_\mathrm{al}$, respectively)
are estimated from the above multiple processes as
\begin{align}
\frac{1}{\tau_\mathrm{ar}} &= \frac{1}{\tau_\mathrm{ar}^\mathrm{UV}}+
\frac{1}{\tau_\mathrm{ar}^\mathrm{SN}},\\
\frac{1}{\tau_\mathrm{al}} &= \frac{1}{\tau_\mathrm{al}^\mathrm{H}}+
\frac{1}{\tau_\mathrm{al}^\mathrm{acc}}.
\end{align}

Since aromatization is predominantly caused by UV irradiation, we assume
that it occurs in the diffuse ISM, where UV radiation can penetrate easily.
On the other hand, aliphatization takes place in a dense region where hydrogen and metals
are easily attached on the dust surface.
Thus, we assume that aliphatization occurs only in the dense ISM.

Since aromatization and aliphatization occur in only one of the two ISM phases,
their time-scales cannot be shorter than the mass exchange time-scales
of the two phases. Here we introduce the time-scale of phase transition from
the dense to diffuse (diffuse to dense) phases as $\tau_\mathrm{12}$ ($\tau_\mathrm{21}$).
Aromatization does not occur more quickly than the supply of the diffuse phase, which
takes place in $\tau_\mathrm{12}$, while aliphatization cannot be faster than
$\tau_\mathrm{21}$.
Thus, the rates of aromatization and aliphatization (denoted as $R_\mathrm{ar}$ and
$R_\mathrm{al}$, respectively) are described as
\begin{align}
R_\mathrm{ar} &= \min\left(\frac{1}{\tau_\mathrm{ar}},\,\frac{1}{\tau_\mathrm{12}}\right) ,\\
R_\mathrm{al} &= \min\left(\frac{1}{\tau_\mathrm{al}},\,\frac{1}{\tau_\mathrm{21}}\right) .
\end{align}
Using the above two rates, we write the time evolution of aromatic fraction as
\begin{align}
\frac{\upartial f_\mathrm{ar}}{\upartial t}=R_\mathrm{ar}(1-f_\mathrm{ar})
-R_\mathrm{al}f_\mathrm{ar}.\label{eq:arom_time}
\end{align}
The phase-transition time-scales are not independent of $\eta_\mathrm{dense}$; here we assume
an equilibrium: $(1-\eta_\mathrm{dense})/{\tau_{21}}={\eta_\mathrm{dense}}/{\tau_{12}}$, obtaining
\begin{align}
\tau_\mathrm{21}=\frac{1-\eta_\mathrm{dense}}{\eta_\mathrm{dense}}\,{\tau_{12}}.\label{eq:phase}
\end{align}
Note also that $\eta_\mathrm{dense}$ is a given parameter in our
one-zone formulation (Section \ref{subsec:size}). We further need to give either $\tau_\mathrm{12}$ or
$\tau_\mathrm{21}$.
The lifetime of dense clouds is likely to be on the order of $10^7$~yr
\citep{McKee:1989aa,Leisawitz:1989aa,Bergin:2007aa,Fukui:2010aa}.
Thus, we simply fix $\tau_{12}=10^7$ yr and move $\tau_{21}$ to satisfy
equation (\ref{eq:phase}).
This is typically longer than the aromatization and aliphatization time-scales, but
much shorter than the chemical enrichment time-scale.
As long as the phase exchange time-scale is between the aromatization and apliphatization
time-scales and the chemical enrichment time-scale,
the aromatic fraction is governed by $\eta_\mathrm{dense}$ (or the ratio of
$\tau_{12}$ to $\tau_{21}$) and is in equilibrium in
most of the chemical enrichment history of the galaxy.
Thus, our results are robust against the detailed choice of the $\tau_\mathrm{12}$ value.

Using the obtained aromatic fraction, we can write the grain mass distributions of
aromatic and non-aromatic carbonaceous grains, $\rho_\mathrm{ar}(m,\, t)$ and
$\rho_\mathrm{non-ar}(m,\, t)$, respectively, as
\begin{align}
&\rho_\mathrm{ar}(m,\, t) = f_\mathrm{ar}(m,\, t)\rho_\mathrm{car}(a,\, t),\\
&\rho_\mathrm{non-ar}(m,\, t) = [1-f_\mathrm{ar}(m,\, t)]\rho_\mathrm{car}(a,\, t).
\end{align}

We assume the minimum and maximum of grain radii for PAHs to be
$a_\mathrm{min,PAH}=a_\mathrm{min}$ and $a_\mathrm{max,PAH}=0.005~\micron$,
respectively \citep{Li:2001aa}.
We calculate the PAH mass density denoted as $\rho_\mathrm{PAH}$ by the following
equation:
\begin{align}
\rho_\mathrm{PAH}(t)=[1-f_\mathrm{sil}(t)]\int_{a_\mathrm{min,PAH}}^{a_\mathrm{max,PAH}}
f_\mathrm{ar}(a,\, t)\rho_\mathrm{dust}(a,\, t)\,\mathrm{d}a.
\end{align}
The PAH-to-gas ratio is defined as
$\mathcal{D}_\mathrm{PAH}\equiv\rho_\mathrm{PAH}/\rho_\mathrm{gas}$, or
\begin{align}
\mathcal{D}_\mathrm{PAH}=[1-f_\mathrm{sil}(t)]\int_{m_\mathrm{min,PAH}}^{m_\mathrm{max,PAH}}
f_\mathrm{ar}(m,\, t)\tilde{\rho}_\mathrm{dust}(m,\, t)\,\mathrm{d}m,
\end{align}
where $m_\mathrm{min/max,PAH}\equiv\frac{4}{3}\upi sa_\mathrm{min/max,PAH}^3$.

\subsection{Calculation of extinction curves}\label{subsec:ext_method}

As an observable quantity, we calculate the extinction curve.
The extinction at wavelength $\lambda $ in units of magnitude ($A_{\lambda }$)
is calculated as
\begin{align} 
A_{\lambda}=(2.5\log_{10} \mathrm{e})L\displaystyle\sum_{i}\displaystyle\int_{0}^{\infty}
n_{i}(a)\,\upi a^{2}Q_{\rm ext}(a, \lambda),
\end{align}
where the subscript $i$ indicates the grain composition (we consider
silicate, aromatic carbon, and non-aromatic carbon), $L$ is the path length,
and $Q_{\rm ext}(a, \lambda)$ is the extinction efficiency factor, which is evaluated
by using the Mie theory \citep[][]{Bohren:1983aa}.
We use the same optical constant of astronomical silicate as adopted
by \citet{Weingartner:2001aa} for silicate, while we adopt graphite in the same paper for
aromatic carbonaceous grains.
Here, graphite is used for a carbonaceous material with
organized atomic structures. Indeed, \citet{Weingartner:2001aa} smoothly connected graphite to PAHs
around $a\sim 20$--50 \AA, and their optical properties of PAHs and graphite are
similar. Moreover, such extremely small grains does not contribute to the
UV--optical extinction significantly.
Therefore, we simply use the optical constants of graphite for all the aromatic grain population.
For non-aromatic species, considering irregular atomic structures,
we adopt the optical constants of amorphous carbon
taken from \citet{Zubko:1996aa} (their ACAR)
\citep[see also][]{Nozawa:2015aa,Hou:2016aa}.
To examine the importance of assumed optical properties for carbonaceous dust,
we also use HAC grains with aromatic and aliphatic structures for an additional
test (Section \ref{subsec:HAC}).

In Section \ref{subsec:separation}, we obtained the grain mass distributions of the
relevant species:
$\rho_\mathrm{sil}(m,\, t)$, $\rho_\mathrm{ar}(m,\, t)$,
and $\rho_\mathrm{non-ar}(m,\, t)$.
Using these mass distributions, the grain size distribution can be calculated by
$n_i(a,\, t)=3\rho_i(m,\, t)/a$ (see equation \ref{eq:rho_vs_n}).
To concentrate on the extinction curve shape, the extinction is normalized to the value in the $V$ band
($\lambda ^{-1}=1.8\,\mu {\rm m}^{-1}$); that is, we output $A_\lambda /A_V$.
In this way, $L$, which is an unknown factor, is cancelled out.


\subsection{Variation of parameters}

Our dust evolution model was already calibrated by observational data of nearby galaxies
(HA19; \citealt{Aoyama:2020aa}). Therefore, we could practically fix the dust evolution model.
However, the one-zone treatment needs to assume the dense gas fraction
($\eta_\mathrm{dense}$)
and the star formation time-scale ($\tau_\mathrm{SF}$), because we do not have
the density information of the ISM.
We adopt $\eta_\mathrm{dense}=0.5$ and $\tau_\mathrm{SF}=5$~Gyr for the
fiducial values, since similar values are also taken by other models for nearby
galaxies (\citealt{Asano:2013aa}; HA19). We also examine $\eta_\mathrm{dense}=0.1$
and 0.9, and $\tau_\mathrm{SF}=0.5$ and 50~Gyr to investigate the effects of these
parameters. $M_\mathrm{g,0}$ is not important for this paper because we
mainly focus on the abundance
indicators (e.g.\ dust-to-gas ratio), for which the total mass is cancelled out. 

\section{Results}\label{sec:result}

Since HA19 and R19 already discussed the evolution of grain size distribution
and the effects of aromatization,
we focus on our new features.
The main new results in this paper concern the PAH abundance and extinction curves;
in particular, we included the time evolution of silicate fraction, and
the effects of aliphatization. Therefore, we describe our results on
the silicate fraction and the aromatic fraction before comparing our results with
observational data.

\subsection{Silicate fraction}

\begin{figure}
\includegraphics[width=0.45\textwidth]{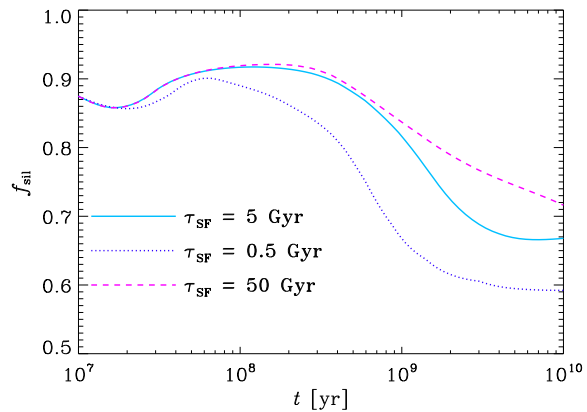}
\caption{Silicate fraction ($f_\mathrm{sil}$) as a function of age.
The solid, dotted, and dashed lines show the results for $\tau_\mathrm{SF}=5$ (fiducial),
0.5, and 50 Gyr, respectively.
\label{fig:fsil}}
\end{figure}

In our model, the silicate fraction, $f_\mathrm{sil}$, is calculated based on the chemical evolution model
(equation \ref{eq:sil}). The evolution of $f_\mathrm{sil}$ depends on the star formation time-scale
$\tau_\mathrm{SF}$. In Fig.\ \ref{fig:fsil}, we show $f_\mathrm{sil}$ as a function of age.
Overall, the silicate fraction is as high as $\sim 0.9$ in the early phase of galaxy evolution,
reflecting the SN yield, while it
drops down to $f_\mathrm{sil}\sim 0.6$--0.7 at later ages
because of the contribution from
the carbon production of low-mass
AGB stars.
The drop of the silicate fraction occurs earlier and more rapidly for
shorter $\tau_\mathrm{SF}$, since the contribution from SNe, which tends to
keep $f_\mathrm{sil}$ high, declines more rapidly.

\subsection{Aromatic fraction and grain size distribution}

\begin{figure}
\includegraphics[width=0.45\textwidth]{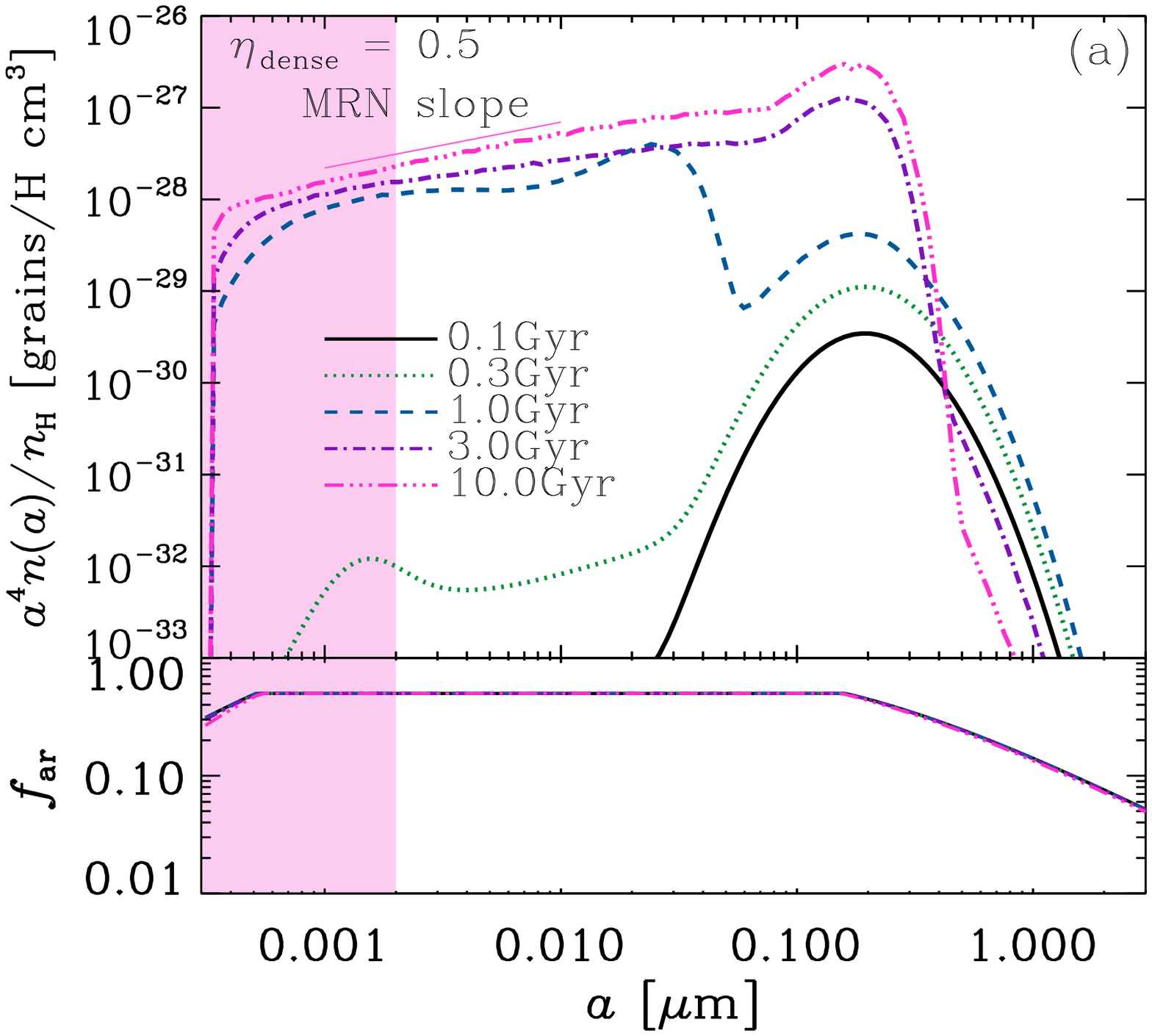}
\includegraphics[width=0.45\textwidth]{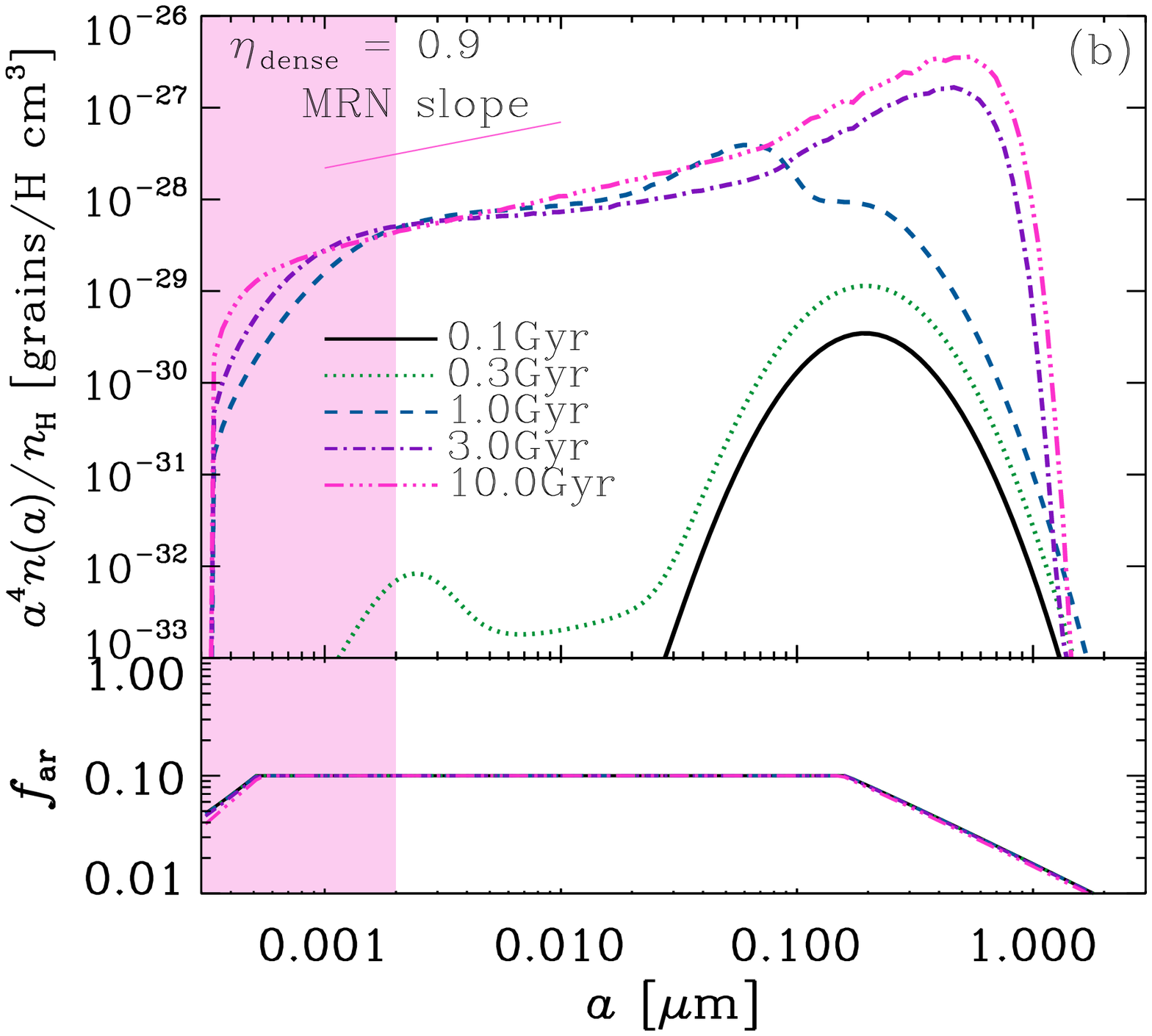}
\includegraphics[width=0.45\textwidth]{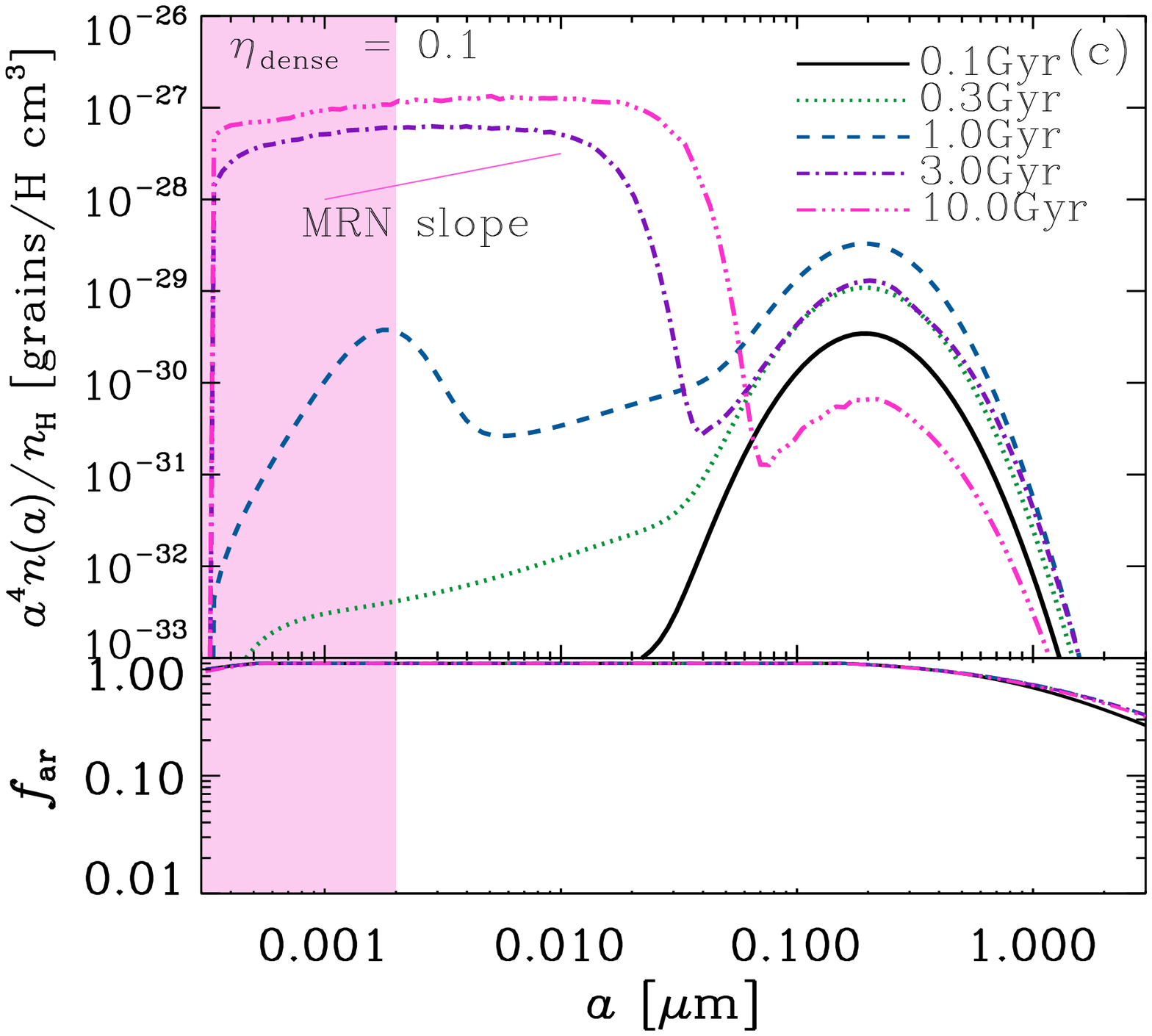}
\caption{Grain size distribution (upper window) and aromatic fraction (lower window)
as a function of grain radius ($a$).
The grain size distribution is multiplied by $a^4$ and divided by $n_\mathrm{H}$,
so that the resulting quantity is proportional to the grain abundance  per $\log a$ relative to the
gas mass.
The solid, dotted, dashed, dot--dashed, and triple-dot--dashed lines show the results
at $t=0.1$, 0.3, 1, 3, and 10~Gyr, respectively. All the lines overlap in the lower
window. The thin dotted straight line in the upper window shows the slope of the MRN
grain size distribution ($n\propto a^{-3.5}$).
{The pink shaded region shows the radius range of PAHs.}
Panels (a), (b), and (c) show the results for $\eta_\mathrm{dense}=0.5$ (fiducial), 0.9, and 0.1,
respectively.
\label{fig:size_far}}
\end{figure}

We present the aromatic fraction ($f_\mathrm{ar}$) together with
the grain size distribution at four representative ages ($t=0.3$, 1, 3, and 10 Gyr)
for various values of $\eta_\mathrm{dense}$ with $\tau_\mathrm{SF}=5$ Gyr in Fig.\ \ref{fig:size_far}.
We first focus on the fiducial case ($\eta_\mathrm{dense}=0.5$; Fig.\ \ref{fig:size_far}a).
The evolution of grain size distribution was already discussed by
HA19, so we only give a brief summary. The grain size distribution is dominated by
large ($a\gtrsim 0.1~\micron$) grains at $t\lesssim 0.3$~Gyr,
reflecting the grain size distribution of stellar sources.
Later, small grains increase by shattering and accretion. In particular,
the drastic increase of small grains between $t=0.3$ and 1~Gyr is driven by accretion,
which is efficient for small grains because of their large surface-to-volume ratios.
This creates a bump in the grain size distribution at $a\sim 0.001$--0.03 $\micron$.
After $t=3$ Gyr, coagulation converts small grains to large ones, smoothing the bump
created by accretion. As a consequence, the grain size distribution converges to a smooth
power-law-like shape similar to the so-called MRN distribution, which is derived
for the Milky Way dust ($n\propto a^{-3.5}$; \citealt{Mathis:1977aa}).

The aromatic fraction changes little as a function of time. This is because
aromatization and aliphatization occur much faster than the chemical
enrichment. In other words, the aromatic fraction is determined by the equilibrium
between aromatization and aliphatization (i.e.\ $\mathrm{d}f_\mathrm{ar}/\mathrm{d}t=0$
in equation \ref{eq:arom_time}). In most of the grain radius range,
$R_\mathrm{ar}=1/\tau_{12}$ and $R_\mathrm{al}=1/\tau_\mathrm{21}$. In this case,
$f_\mathrm{ar}=1-\eta_\mathrm{dense}$ in the equilibrium condition, noting the
relation between $\tau_{12}$ and $\tau_{21}$ (equation \ref{eq:phase}).
In the fiducial case, we indeed observe that $f_\mathrm{ar}=0.5$
at $a\sim 0.001$--0.1 $\micron$. For smaller and larger $a$,
the aromatization time-scale becomes longer than $\tau_\mathrm{12}$ or $\tau_{21}$,
so that $f_\mathrm{ar}$ is lower (but the equilibrium still holds).

Since $\eta_\mathrm{dense}$ is important for the aromatic fraction, we also
show the results for $\eta_\mathrm{dense}=0.1$ and 0.9 in Fig.\ \ref{fig:size_far}.
The grain size distribution also depends on $\eta_\mathrm{dense}$.
In the early ($t<0.3$ Gyr) phase, the grain size distribution is not sensitive to the
change of $\eta_\mathrm{dense}$, since the dust abundance is dominated by
stellar dust production (not by interstellar processing).
At $t=0.3$ Gyr, the creation of small grains by shattering is the most efficient for
the smallest $\eta_\mathrm{dense}=0.1$, since shattering occurs in the diffuse ISM.
Since the subsequent enhancement of small grains is driven by accretion, which occurs
in the dense ISM, the small-grain abundance is higher for $\eta_\mathrm{dense}=0.9$
than for $\eta_\mathrm{dense}=0.1$ at $t\gtrsim 1$ Gyr.
In the case of the highest $\eta_\mathrm{dense}=0.9$, coagulation is also efficient at
$t\gtrsim 1$~Gyr, suppressing the abundance of the smallest grains.
At $t\gtrsim 3$~Gyr, the large-grain abundance at $a\gtrsim 0.1~\micron$ is
very sensitive to $\eta_\mathrm{dense}$
because the large-grain formation by coagulation occurs more efficiently
for larger $\eta_\mathrm{dense}$. Therefore, the difference in $\eta_\mathrm{dense}$ has a dramatic
impact on the grain size distribution in later epochs.

As discussed above, the aromatic fraction is determined by the equilibrium value
$f_\mathrm{ar}=1-\eta_\mathrm{dense}$ in most of the grain radius range.
Thus, the carbonaceous dust is more aromatized if the diffuse gas is dominant.
This means that the PAH abundance is expected to be higher for
lower $\eta_\mathrm{dense}$. We will see this again in Section \ref{subsec:abundance}.

\subsection{Extinction curves}\label{subsec:ext_result}

\begin{figure}
\includegraphics[width=0.45\textwidth]{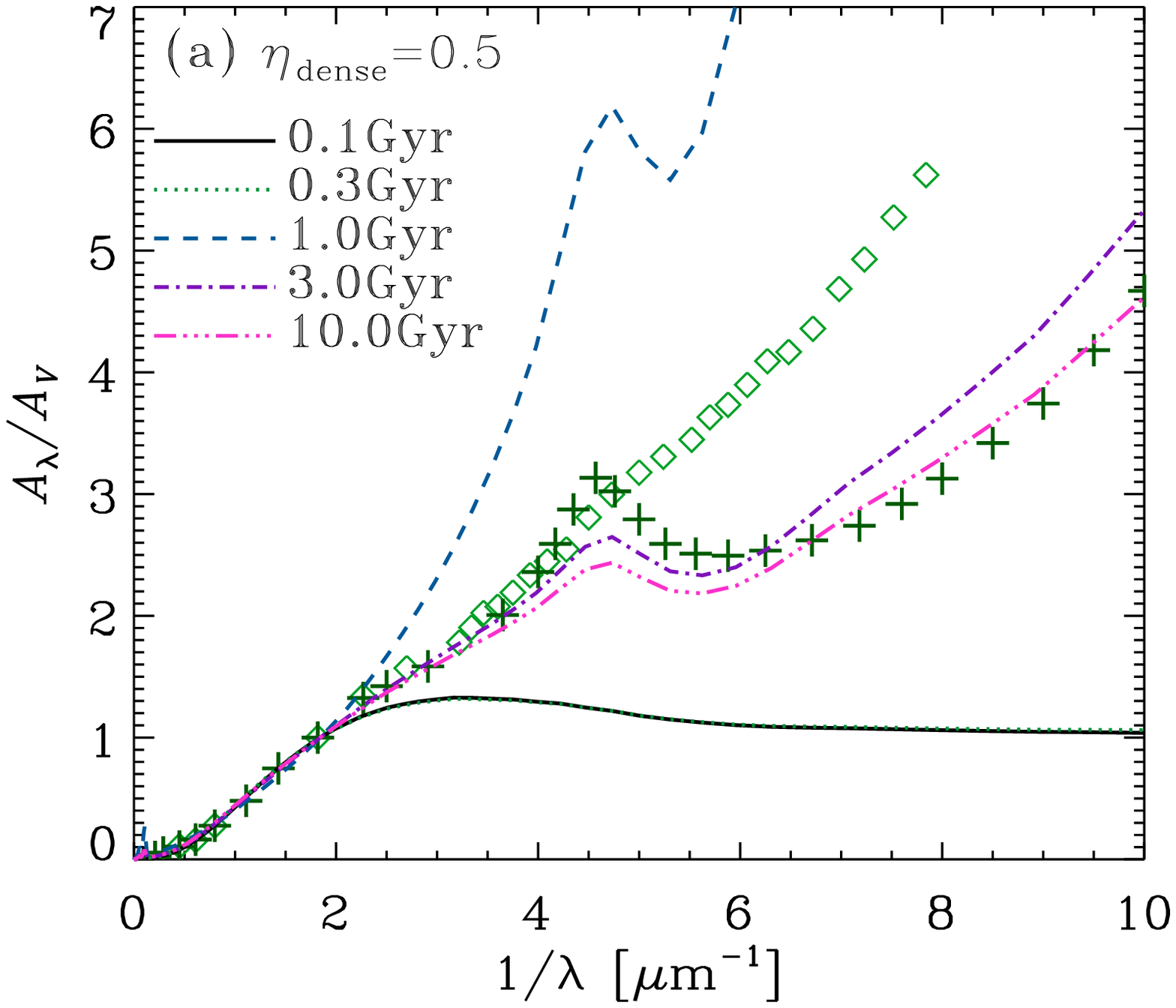}
\includegraphics[width=0.45\textwidth]{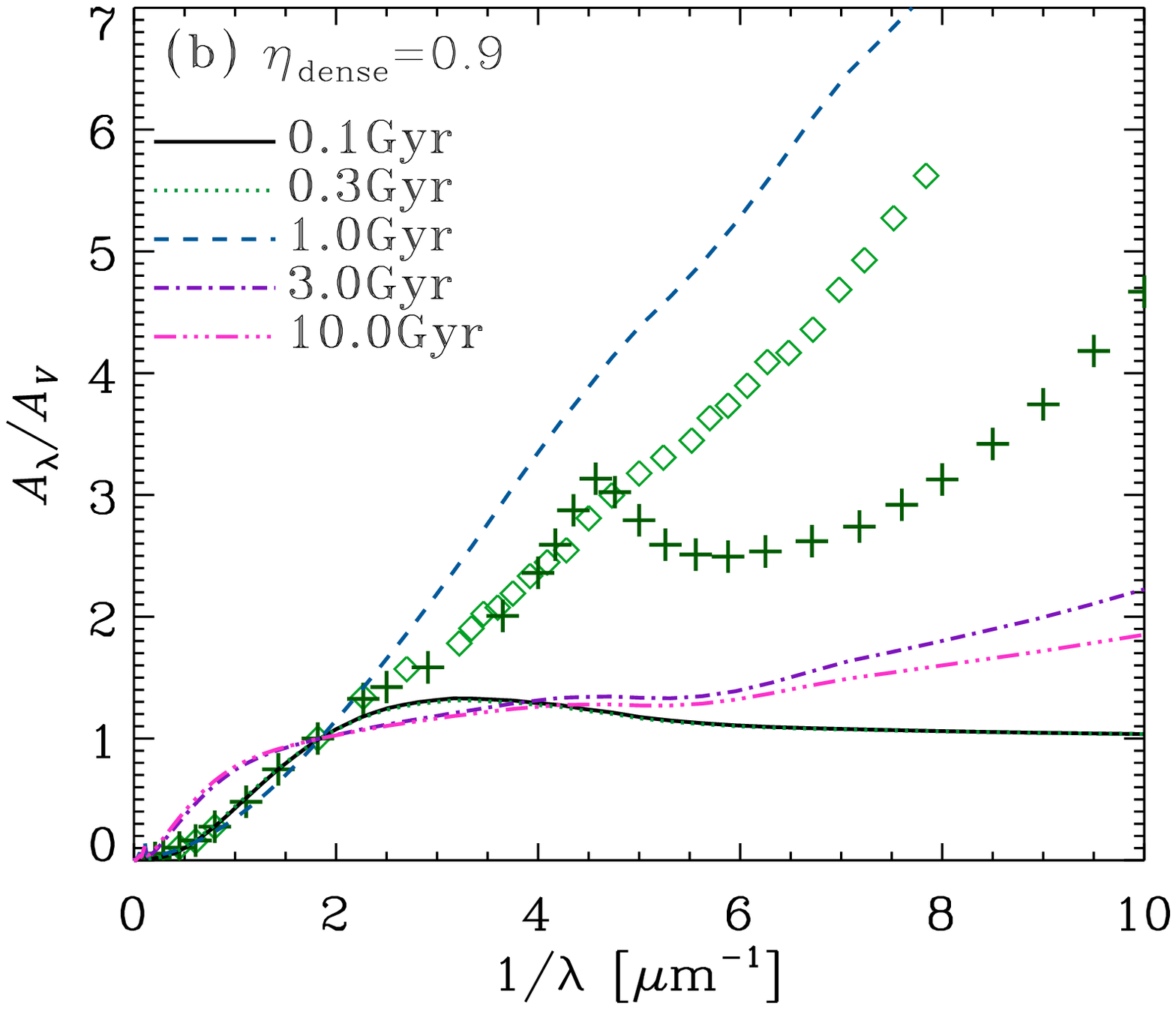}
\includegraphics[width=0.45\textwidth]{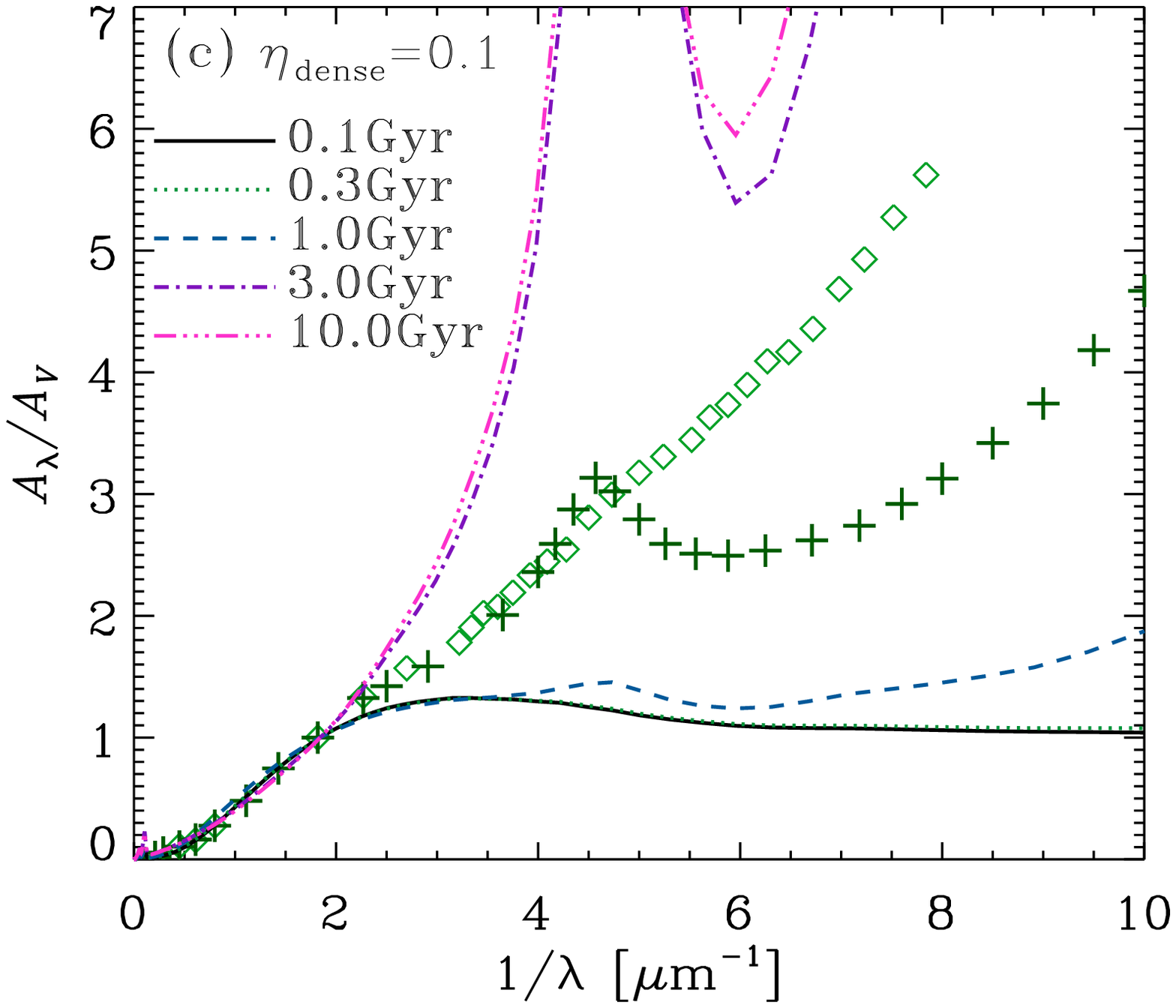}
\caption{
Extinction curves corresponding to the three cases for
$\eta_\mathrm{dense}$ (= 0.5, 0.9, and 0.1 for panels a, b,
and c, respectively) shown
in Fig.\ \ref{fig:size_far}. The solid, dotted, dashed, dot--dashed,
and triple-dot--dashed lines show the results at
$t=0.1$, 0.3, 1, 3, and 10 Gyr, respectively.
The lines at $t=0.1$ and 0.3 Gyr are
indistinguishable. The crosses and diamonds show the
observational data of the Milky Way and SMC extinction
curves, respectively, taken from \citet{Pei:1992aa}.
\label{fig:ext}}
\end{figure}

Based on the grain size distributions shown above, we calculate the extinction
curves by the method explained in Section \ref{subsec:ext_method}.
In Fig.~\ref{fig:ext}, we present the extinction curves corresponding to the
grain size distributions shown in Fig.\ \ref{fig:size_far}.
As expected, the extinction curve is flat in the early epoch
($t\lesssim 0.3$ Gyr) since most of the grains are large.
For $\eta_\mathrm{dense}=0.5$ and 0.9, the extinction curve becomes
the steepest at $t\sim 1$ Gyr because accretion drastically enhances the
small-grain abundance. For $\eta_\mathrm{dense}=0.1$, the extinction curve
is still flat at $t=1$~Gyr because accretion, which only occurs in the dense ISM,
is not efficient. At $t>1$ Gyr, the extinction curve flattens for
$\eta_\mathrm{dense}=0.5$ and 0.9 because of coagulation, while it
remains very steep for $\eta_\mathrm{dense}=0.1$ because of inefficient
coagulation.
The strength of the 2175 \AA\ bump created by small graphite grains
correlates with the steepness of the extinction curve.
The above evolutionary trend confirms the evolution of grain size distribution
calculated by one-zone models
\citep{Asano:2014aa,Hou:2016aa} and hydrodynamic simulations
\citep{Hou:2017aa,Hou:2019aa}.
Because a significant fraction of carbonaceous grains are aromatized for
$\eta_\mathrm{dense}=0.5$ and 0.1, the 2175 \AA\ bump is prominent.
For $\eta_\mathrm{dense}=0.9$, the aromatic fraction is low ($\sim 0.1$)
so that the extinction curves are practically bumpless.
Thus, a large $\eta_\mathrm{dense}$ is not suitable for explaining the Milky Way
extinction curves (or extinction curves with a bump in general).

We emphasize that the extinction curve in the fiducial model becomes similar
to the Milky Way extinction curve at $t\gtrsim 3$ Gyr.
The extinction curves at $t\gtrsim 3$ Gyr are steeper than
the Milky Way curve with too
strong a bump for $\eta_\mathrm{dense}=0.1$, while they are
much flatter than the SMC curve for
$\eta_\mathrm{dense}=0.9$.
It is also worth noting that the extinction curve at $t=1$ Gyr
for $\eta_\mathrm{dense}=0.9$ is steep without a prominent bump. This
implies a possibility of reproducing the SMC extinction curve with a high
$\eta_\mathrm{dense}$ (or a low aromatic fraction). We discuss this issue later
in Section \ref{subsec:starburst}.

\subsection{Evolution of PAH abundance}\label{subsec:abundance}

\begin{figure*}
\includegraphics[width=0.48\textwidth]{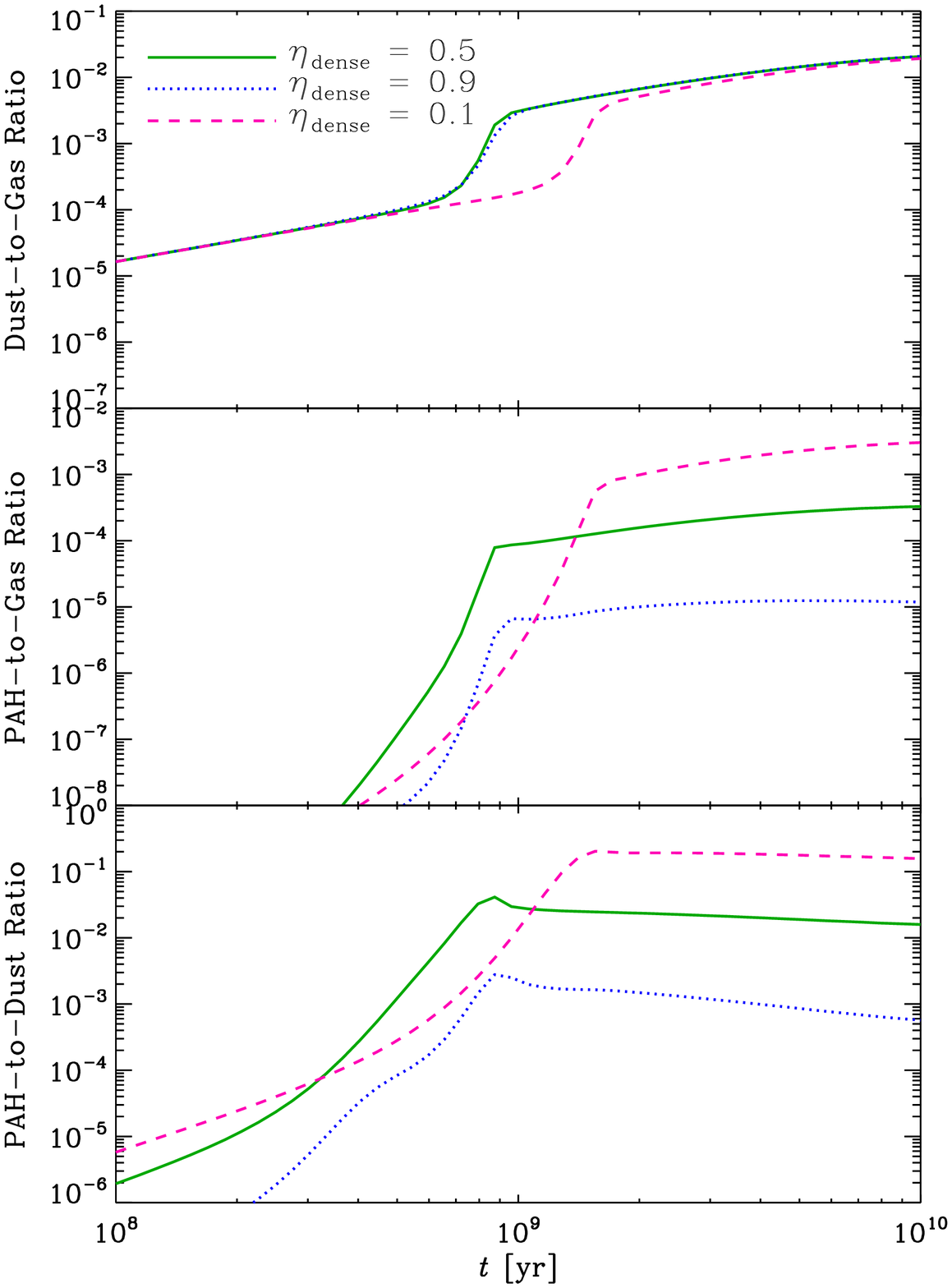}
\includegraphics[width=0.48\textwidth]{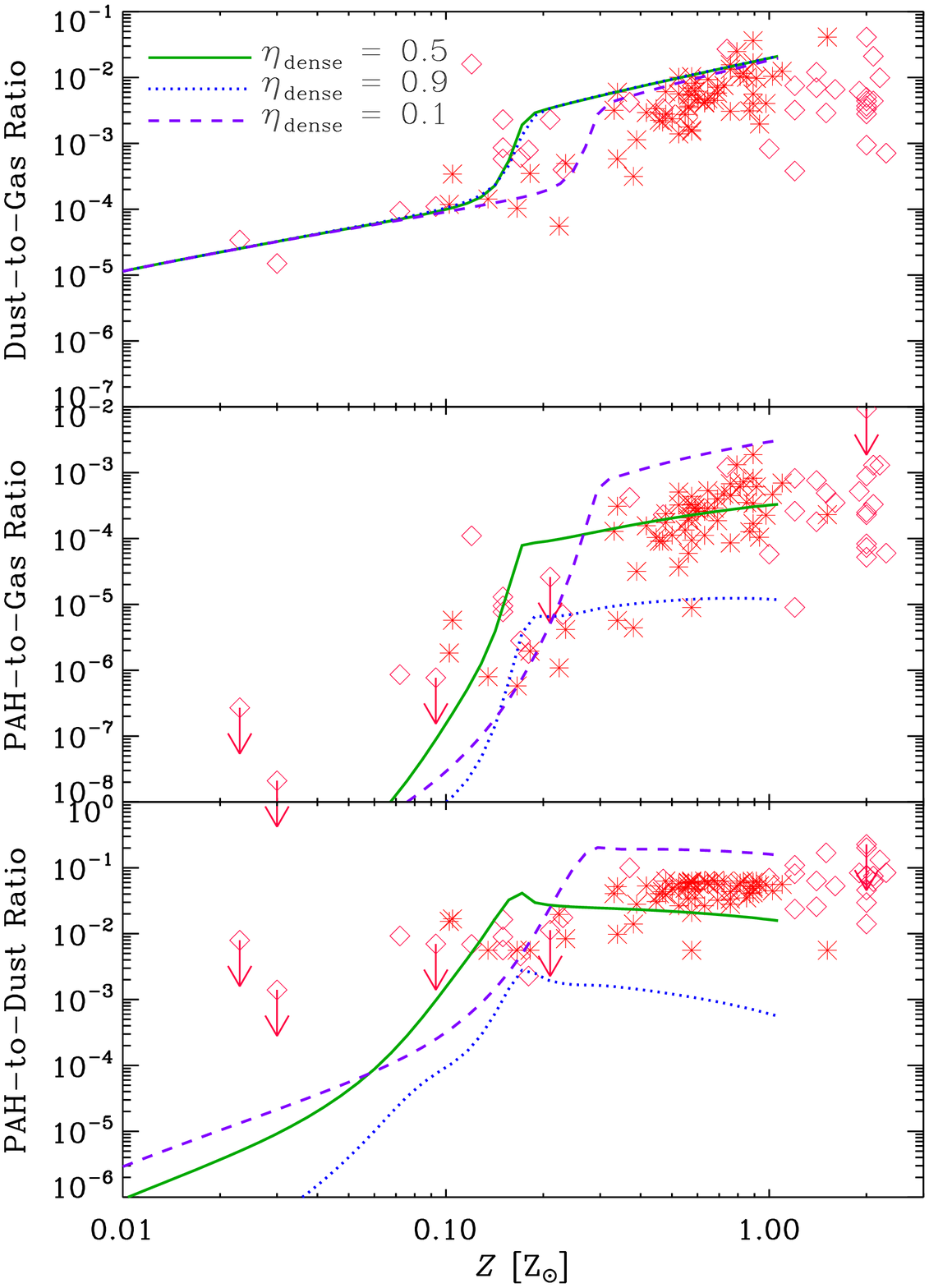}
\caption{Evolution of dust-to-gas ratio (top), {PAH-to-gas} ratio (middle),
and PAH-to-dust ratio (bottom). The solid, dotted, and dashed lines
show the results for $\eta_\mathrm{dense}=0.5$ (fiducial), 0.9, and 0.1,
respectively. The evolution is shown as a function of age (left) and
metallicity (right). In the right panel, observational data are also shown:
the diamonds and asterisks present the data taken from
\citet{Galliano:2008aa} and \citet{Draine:2007ab}, respectively, with
arrows showing upper limits.
\label{fig:abundance}}
\end{figure*}

Now we examine the evolution of PAH abundance.
For the interpretation of the PAH abundance, the total dust abundance
(dust-to-gas ratio, $\mathcal{D}$) is of fundamental importance since
PAHs form as a result of dust processing. Thus, we show the evolution of
dust-to-gas ratio ($\mathcal{D}$),
PAH-to-gas ratio ($\mathcal{D}_\mathrm{PAH}$),
and PAH-to-dust ratio ($\mathcal{D}_\mathrm{PAH}/\mathcal{D}$) in
Fig.~\ref{fig:abundance}
for the three values of $\eta_\mathrm{dense}$ with $\tau_\mathrm{SF}=5$~Gyr.
Since the dust and PAH abundances are extremely poor at $t<10^8$~yr,
we only show the results at $t>10^8$ yr, noting that the behaviour at $t<10^8$ yr
is easily inferred from the extrapolation of that at older ages.

In Fig.\ \ref{fig:abundance}, we observe an increase of dust-to-gas ratio
throughout the entire age range.
The rapid increase around $t\sim 10^9$ yr for $\eta_\mathrm{dense}=0.5$ and 0.9
is due to dust growth by accretion. The dust growth occurs later for
$\eta_\mathrm{dense}=0.1$ because of a smaller fraction of the dense ISM hosting
accretion. Since accretion drastically increases the small-grain abundance
as shown above,  it also induces the rapid increase of PAH abundance
at $t\sim 1$--$2\times 10^8$ yr.
Note that accretion does not directly produce PAHs but form
non-aromatic carbonaceous grains, a part of which are quickly aromatized.
Moreover, shattering also helps to convert the aromatic grains with $a>50$ \AA\
to small PAHs. After this rapidly increasing phase,
the PAH-to-gas ratio saturates
because the depletion of small grains by coagulation is balanced with the
new creation of small grains by accretion and shattering.
The final PAH abundance is sensitive to $\eta_\mathrm{dense}$ mainly because the
PAH abundance is strongly suppressed by coagulation for high $\eta_\mathrm{dense}$.

In order to show the relative abundance of PAHs to dust, we show the
PAH-to-dust ratio in the bottom panels of Fig.\ \ref{fig:abundance}.
The increase of small-grain abundance is a nonlinear process in the sense
that it is caused by the collision between grains (shattering)
or a grain and metals (accretion). Thus, PAHs originating from small grains
increase more steeply than the total dust abundance. This is why
the PAH-to-dust ratio rises at $t\lesssim 10^8$ yr.
The increase of PAH-to-dust ratio is saturated afterwards, and it
rather decreases in the case of high $\eta_\mathrm{dense}$ because of coagulation.
Note that coagulation decreases the PAH abundance while it does not
change the entire dust abundance. The PAH-to-dust ratio is higher for
smaller $\eta_\mathrm{dense}$ as expected from the PAH-to-gas ratio.

The age is difficult to determine, so that the metallicity, which is better measured,
is often used to constrain the evolution model of dust content in galaxies
\citep[e.g.][]{Lisenfeld:1998aa}.
In Fig.~\ref{fig:abundance}, we also show the three quantities
as a function of metallicity. For comparison, we present the observational data
of nearby galaxies taken
from \citet{Galliano:2008aa} and \citet{Draine:2007ab}
(used also by \citealt{Seok:2014aa} and R19).
The observational metallicity data are based on the oxygen abundance.
The solar oxygen abundance is uncertain, and we simply followed the solar
abundance values adopted in those original papers [$12+\log (\mathrm{O/H})_{\sun}=8.83$
and $12+\log (\mathrm{O/H})_{\sun}=8.69$ for
\citealt{Galliano:2008aa} and \citealt{Draine:2007ab}, respectively], keeping in mind that there is
a factor 2 uncertainty in the solar metallicity.
The typical error of the PAH abundance is a factor of $\sim 2$.

The observed relation between dust-to-gas ratio and metallicity is reproduced well
at both low and high metallicities. The calculation results tend to overproduce the dust-to-gas
ratio at high metallicity, but they are within the scatter of the observational data.
The theoretical curves strop at $Z\sim 1$~Z$_{\sun}$, corresponding to the metallicity
achieved at $t=10$ Gyr.
The observed PAH-to-gas ratios are also explained by the theoretical models; in particular,
the deficiency at $Z<0.1$ Z$_{\sun}$ and the richness at $Z>0.1$~Z$_{\sun}$ in
the PAH abundance are observationally suggested by \citet{Draine:2007aa}, and
are reproduced well by our models.
The PAH-to-dust ratio is also consistent with the observational data.
The model with $\eta_\mathrm{dense}=0.9$ underproduces the PAH abundance, which is
due to too strong coagulation. This implies that such a high cold gas fraction is not
applicable to nearby galaxies.

The very steep increase of the PAH abundance at $Z\sim 0.1$~Z$_{\sun}$ is
consistent with \citet{Seok:2014aa}'s result. We observe more gradual increase in R19,
who used the information of individual fluid elements within a galaxy.
Our one-zone model, by treatment, predicts a coherent evolution in the entire
galaxy; thus, there is a certain metallicity ($\sim$0.1 Z$_{\sun}$ in our
case), where the increase of dust mass by accretion becomes prominent. In contrast, R19 considered
different evolutionary paths in the gas density among individual fluid elements; thus,
the rapid increase of dust mass does not occur coherently at a certain metallicity.
However, we emphasize that a nonlinear increase of PAH abundance is commonly
seen between our model and R19's. Therefore, we confirm that the PAH abundance
is more sensitive to the metallicity than the total dust abundance is.

\subsection{Effects of star formation history}\label{subsec:sfh}

\begin{figure}
\begin{center}
\includegraphics[width=0.45\textwidth]{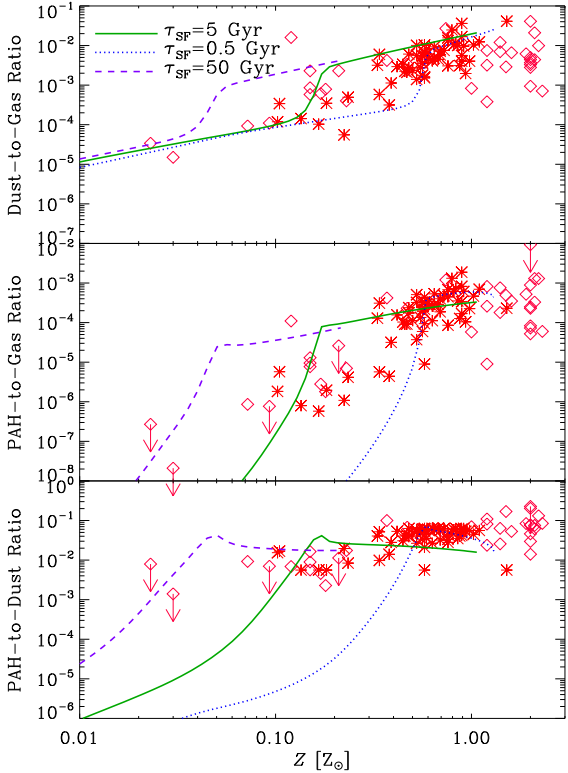}
\end{center}
\caption{Same as Fig.\ \ref{fig:abundance} but for various star formation
time-scales ($\tau_\mathrm{SF}$) with a fixed $\eta_\mathrm{dense}=0.5$.
The solid, dotted, and dashed lines
show the results for $\tau_\mathrm{SF}=5$ (fiducial), 0.5, and 50 Gyr, respectively.
\label{fig:abundance_tau}}
\end{figure}

Various types of galaxies in a wide range of redshift show a large
variety of star formation histories. For simplicity, we represent the variation
of star formation history by the change of $\tau_\mathrm{SF}$.
We examine $\tau_\mathrm{SF}=0.5$, 5, and 50~Gyr with a fixed
$\eta_\mathrm{dense}=0.5$.
In Fig.\ \ref{fig:abundance_tau}, we show the evolution of the dust and PAH abundances
for the three cases of $\tau_\mathrm{SF}$.
For $\tau_\mathrm{SF}=0.5$~Gyr, we stop the calculation at $t=3$ Gyr, when the
star formation declines and no further metal enrichment occurs.

We observe in Fig.\ \ref{fig:abundance_tau} that the metallicity at which the
dust-to-gas ratio rises steeply is sensitive to $\tau_\mathrm{SF}$.
As already found by \citet{Asano:2013ab}, the metallicity at the rapid rise of
dust-to-gas ratio is proportional to $\tau_\mathrm{SF}^{-1/2}$.
Compared with the nearby galaxy data, long $\tau_\mathrm{SF}$ (such as 50 Gyr) give
too rich a dust content at $Z<0.1$~Z$_{\sun}$.
In the case of $\tau_\mathrm{SF}=50$ Gyr, the lines end at
$\sim$0.2 Z$_{\sun}$, corresponding to the metallicity reached at $t=10$ Gyr.
The PAH abundances are also consistent with the observational data for
$\tau_\mathrm{SF}=5$ and 0.5 Gyr. If the star formation time-scales in the
nearby galaxies lie between 5 and 0.5 Gyr, most of the data points at
intermediate metallicities ($\sim$0.2--0.5 Z$_{\sun}$) can be explained.
At high metallicity, the PAH abundance converges to the same value
regardless of $\tau_\mathrm{SF}$. Thus, the change of $\tau_\mathrm{SF}$
does not produce a scatter of the PAH abundance at high metallicity.
(Recall that variation of $\eta_\mathrm{dense}$ produces such a scatter
as shown in Fig.\ \ref{fig:abundance}.)
Since the rapid rise of PAH abundance coincides with the increase of
dust-to-gas ratio by accretion, the metallicity at which the system becomes rich in PAHs
scales with the star formation time-scale as $\tau_\mathrm{SF}^{-1/2}$.
Using this scaling, the metallicity level
at which the rapid rise of PAH abundance takes place is roughly estimated as
$0.1(\tau_\mathrm{SF}/5~\mathrm{Gyr})^{1/2}$~Z$_{\sun}$.

\begin{figure}
\begin{center}
\includegraphics[width=0.45\textwidth]{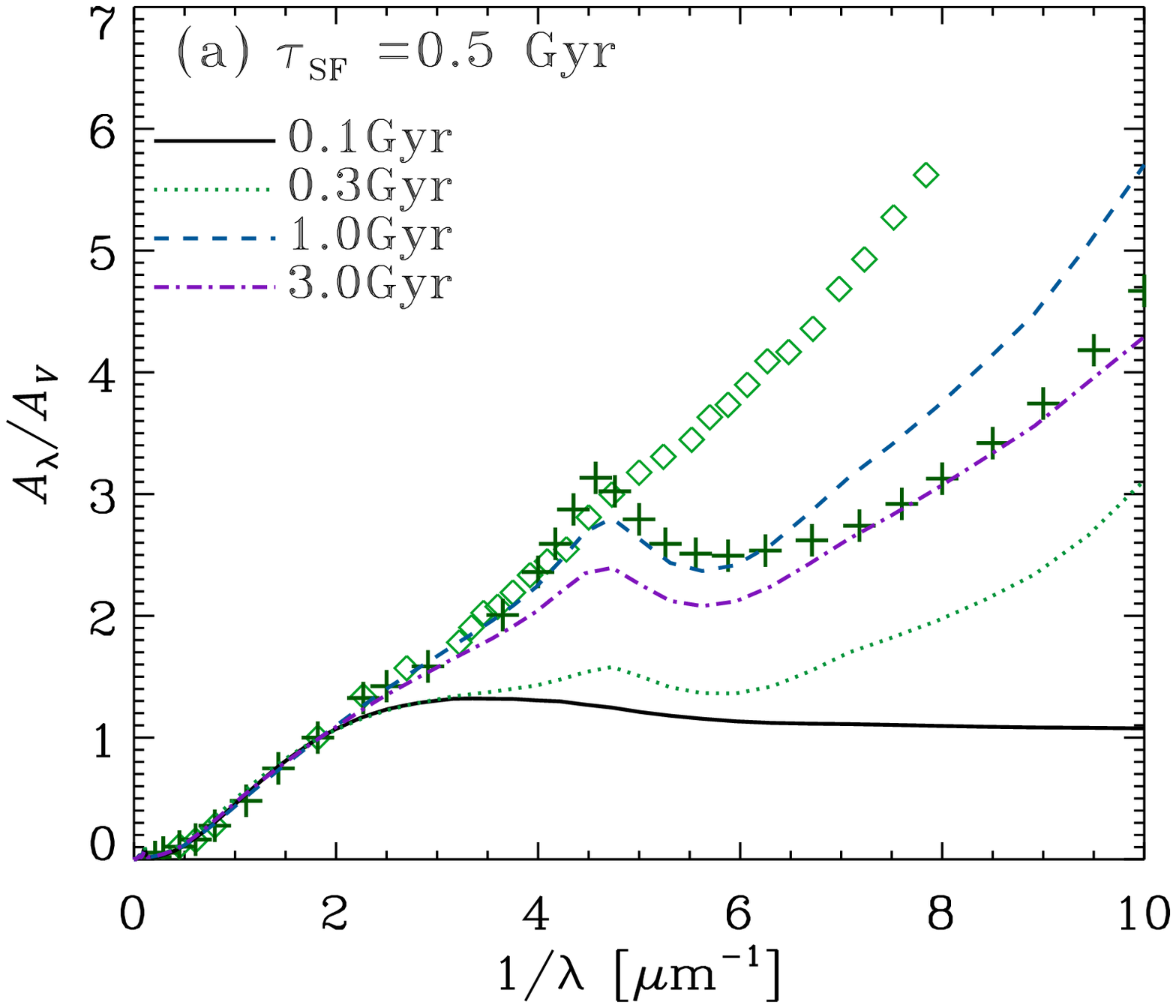}
\includegraphics[width=0.45\textwidth]{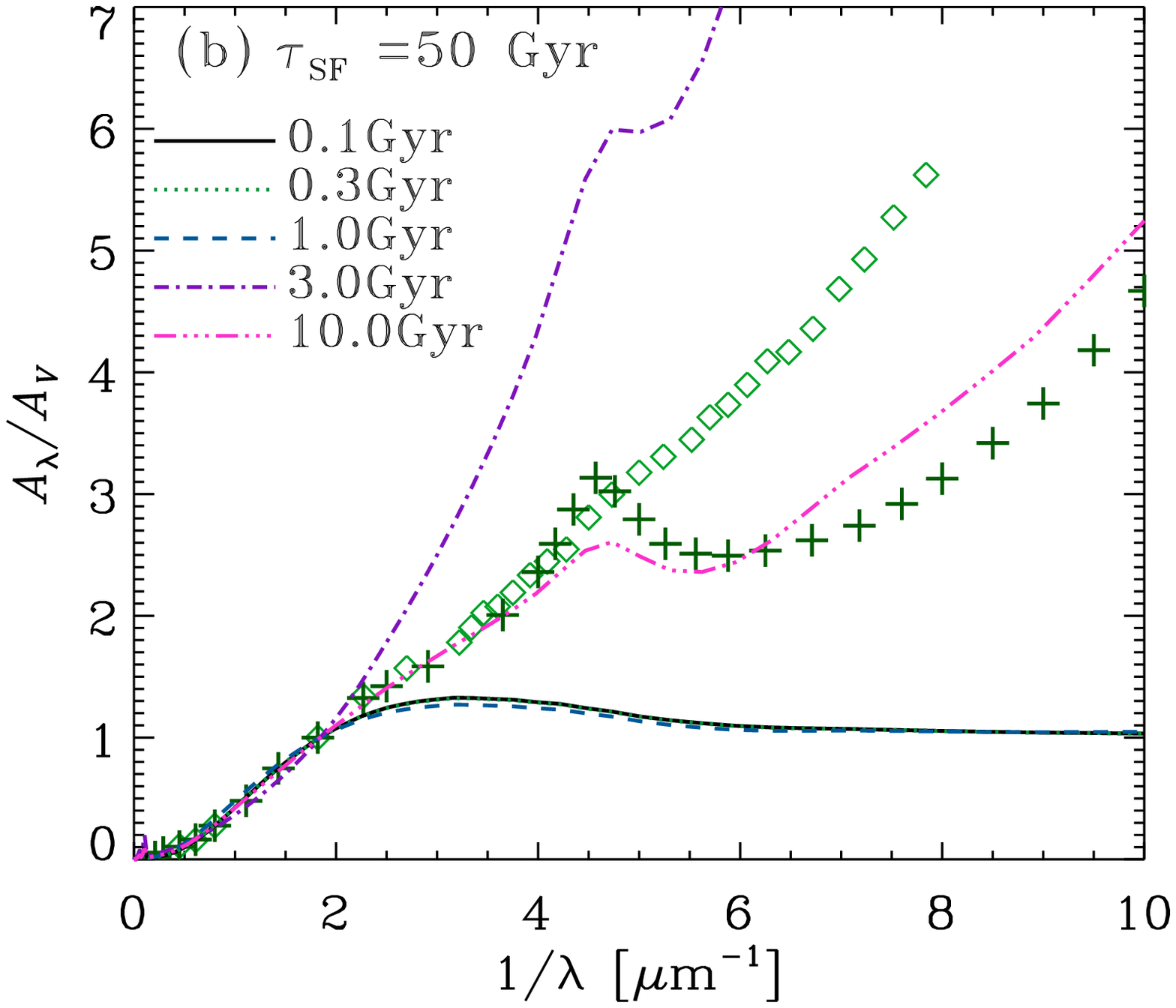}
\end{center}
\caption{Same as Fig.\ \ref{fig:ext} but for different star formation
time-scales ($\tau_\mathrm{SF}$) with $\eta_\mathrm{dense}$ fixed
to 0.5. We adopt $\tau_\mathrm{SF}=0.5$ and 50 Gyr in Panels (a) and (b),
respectively. The calculation is stopped at $t=3$ Gyr for $\tau_\mathrm{SF}=0.5$ Gyr.
Note that the results for $\tau_\mathrm{SF}=5$ Gyr is shown in Fig.\ \ref{fig:ext}.
\label{fig:ext_tau}}
\end{figure}

In Fig.~\ref{fig:ext_tau}, we show the evolution of extinction curves for
$\tau_\mathrm{SF}=0.5$ and 50~Gyr with $\eta_\mathrm{dense}=0.5$.
Note that the result for $\tau_\mathrm{SF}=5$~Gyr is shown in Fig.\ \ref{fig:ext}.
We observe that the extinction curves tend to become similar to the Milky Way curve
at old ages. The extinction curve (grain size distribution) evolves
more quickly for shorter $\tau_\mathrm{SF}$ because of faster dust enrichment.
However, the evolutionary sequence of extinction curves is common for all
$\tau_\mathrm{SF}$; that is, the extinction curves are flat at young ages
when the dust production is dominated by stellar sources; they become
drastically steep at an intermediate age when accretion rapidly increases the
small-grain abundance; and subsequently, they become flatter
and converge to a shape similar to the Milky Way curve.
In the case of $\tau_\mathrm{SF}=0.5$~Gyr, the very steep phase does not
appear in the figure since it occurs only in a short period (see also Section \ref{subsec:starburst}).
Because the evolutionary time-scale of grain size distribution is scaled as
$\propto\tau_\mathrm{SF}^{1/2}$ \citep{Asano:2013ab},
the extinction curves are similar for the
same value of $t/\tau_\mathrm{SF}^{1/2}$. For example, the extinction curve
at $t=1$ Gyr for $\tau_\mathrm{SF}=0.5$ Gyr is similar to that at
$t=10$ Gyr for $\tau_\mathrm{SF}=50$ Gyr.

\section{Discussion}\label{sec:discussion}

\subsection{Robustness of grain size distribution}\label{subsec:robustness}

\begin{figure}
\begin{center}
\includegraphics[width=0.45\textwidth]{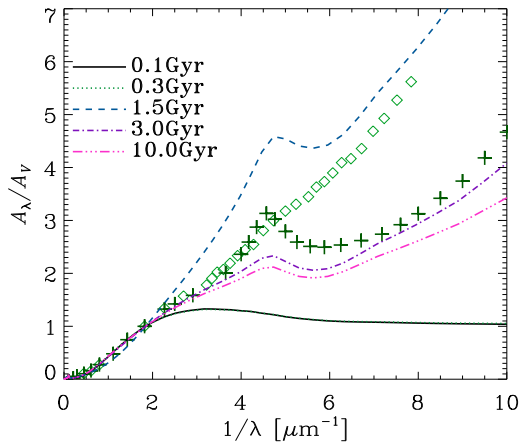}
\end{center}
\caption{Same as Fig.\ \ref{fig:ext} but the calculation of grain size distribution
is based on the silicate material properties. (Note that we still separate the
size distribution into the relevant species in the same way as above; see text).
We adopt the fiducial parameter values ($\eta_\mathrm{dense}=0.5$
and $\tau_\mathrm{SF}=5$ Gyr).
\label{fig:ext_sil}}
\end{figure}

In our model, we assumed a single species (graphite) in computing the
evolution of grain size distribution, which is subsequently separated into
the relevant species. In order to see how robust the results
are against the assumed grain properties, we
here examine the evolution of grain size distribution by assuming the
silicate properties (i.e.\ by using the quantities for silicate
described in Section \ref{subsec:size}).

In Fig.\ \ref{fig:ext_sil}, we show the resulting extinction curves
based on the grain size distributions calculated by the silicate properties.
This figure is to be compared with Fig.\ \ref{fig:ext}a.
Note that the fractions of silicate, graphite, and amorphous carbon
are the same as the above calculations. At $t\lesssim 0.3$ Gyr, the extinction curve is not sensitive
to the change of grain properties since the grain size distribution has the same
functional (lognormal) form. The steepening of extinction curve occurs more slowly
because silicate has a higher value of $Q_\mathrm{D}^\star$ (specific
impact energy that causes catastrophic disruption), leading to less efficient
shattering. Accordingly, the drastic steepening of extinction curve caused by
accretion is delayed; since the extinction curve at $t=1$ Gyr is still flat,
we show the extinction curve at $t=1.5$ Gyr, when it
becomes significantly steep.
Coagulation is rather more efficient if we adopt the silicate properties,
since the higher grain material density ($s$) indicates a higher grain velocity
(HA19). This leads to a higher abundance of grains
at $a\sim 0.1$--0.3~$\micron$ and flatter extinction curves with a less
prominent 2175 \AA\ bump at
$t=3$ and 10 Gyr.

In summary, the material properties indeed affect the grain size distributions
and the effects of adopting silicate instead of graphite are summarized
as following: (i) The increase of small
grains by shattering and accretion is delayed because of less efficient
shattering. Accordingly, the steepening of extinction curve occurs later,
but this delay is much shorter than $\tau_\mathrm{SF}$. Therefore, the
difference in the timing of the steepening of extinction curve is not
important if we consider galaxy evolution on the star-forming time-scale.
(ii) Coagulation is rather more efficient, leading to a flatter extinction curve
at $t\gtrsim 3$ Gyr. This causes a small difference in $A_\lambda /A_V$ (by 30 per cent).
It is also true that the treatment of coagulation depends
on the grain shapes (or fluffiness), which needs further investigations.
Therefore, a possible imprint of grain material properties on the resulting
grain size distribution needs further detailed studies.

\subsection{Other possible optical properties of aromatic carbons}
\label{subsec:HAC}

\begin{figure}
\begin{center}
\includegraphics[width=0.45\textwidth]{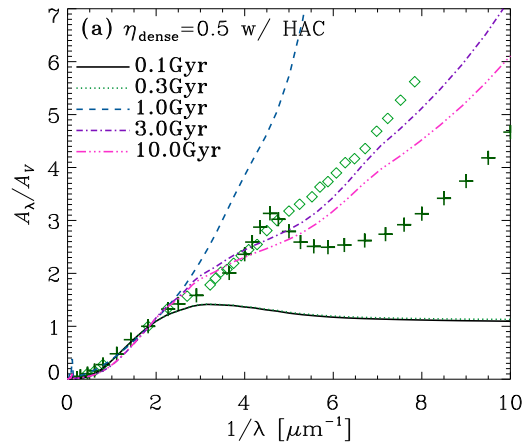}
\includegraphics[width=0.45\textwidth]{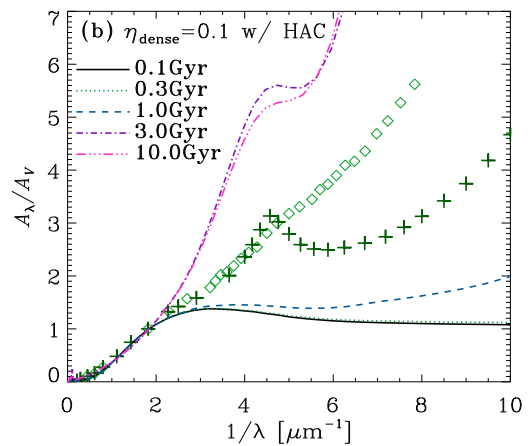}
\end{center}
\caption{Same as Fig.\ \ref{fig:ext} but using the HAC optical properties
for the carbonaceous dust (see text for details). Panels (a) and (b) show
the results for $\eta_\mathrm{dense}=0.5$ and 0.1, respectively, with
$\tau_\mathrm{SF}=5$ Gyr.
\label{fig:ext_HAC}}
\end{figure}

In the above, we have adopted graphite as a representative carbonaceous
species with regular atomic structures (i.e.\ the aromatic component in our
model) in calculating the extinction curve. As shown by \citet{Zubko:2004aa},
the same extinction curve could be
reproduced by different combinations of dust species and grain size distributions.
This means that it is worth examining how the extinction
curves change by the assumed optical properties.

Other than the graphite--silicate mixture, one of the most developed dust models
that explain the Milky Way extinction curve and dust emission is
the one constructed by \citet{Jones:2013aa}. Their dust model is composed of
(hydrogenated) amorphous carbons and amorphous silicate which could have
mantles of carbonaceous materials. However, their model needs a significantly enhanced abundance of
small carbonaceous grains (relative to the MRN grain size distribution)
to reproduce the Milky Way extinction curve because
aromatic hydrocarbons adopted by \citet{Jones:2013aa} have
weaker 2175 \AA\ bump intensity than graphite.
The grain size distributions in our
model, which basically converge to the MRN-like distribution, do not show
such an enhancement of small carbonaceous grains. Thus, if we adopt
\citet{Jones:2013aa}'s optical properties and our grain size distribution,
we predict too weak a 2175 \AA\ bump to reproduce the Milky Way extinction curve.

Nevertheless, it is interesting to investigate what kind of extinction curve
our model produces if we adopt the carbonaceous dust properties similar to
those adopted by \citet{Jones:2013aa}. For this purpose, we adopt the
optical properties of HAC adopted by \citet{Murga:2019aa}:
fully aliphatic grains with $E_\mathrm{g}=2.67$ eV
and fully aromatic grains with $E_\mathrm{g}=0.01$ eV for the aromatic and
non-aromatic components, respectively.
We adopt the neutral case, but we confirmed that adopting the ionized HAC does not
change the extinction curves.
In Fig.\ \ref{fig:ext_HAC}a, we show the evolution of
extinction curve for the fiducial case but using the HAC optical properties
for the fiducial parameter sets ($\eta_\mathrm{dense}=0.5$ and $\tau_\mathrm{SF}=5$ Gyr).
We observe that, as expected above, the 2175 \AA\ bump is not prominent.
The extinction curves at later epochs are rather consistent with the SMC
extinction curve.

One may wonder if an enhancement of small grains as seen in the model of
low $\eta_\mathrm{dense}$ would produce a prominent carbon bump. To answer
this question, we examine the case of $\eta_\mathrm{dense}=0.1$, which shows
an enhancement of small grains (see Fig.\ \ref{fig:size_far}).
We indeed see a bump around 2175~\AA\ at $t\geq 3$ Gyr,
when the abundance of small grains relative to large grains is the most enhanced.
However, the extinction curve stays much steeper than the Milky Way curve because
coagulation does not take place efficiently.
Therefore, it is difficult to reproduce both the strong 2175 \AA\ bump and
the steepness of the Milky Way extinction curve with our model if we adopt the HAC
optical properties. We need both enhancement of small grains and
efficient coagulation to reproduce the Milky Way extinction curve, and these two
requirements are contradictory in our model.

Even in the graphite--silicate model, \citet{Weingartner:2001aa} also
suggested an enhancement of small ($\lesssim 10^{-3}~\micron$) carbonaceous
dust grains, mainly to reproduce the MIR emission in the Milky Way \citep{Li:2001aa}.
Therefore, some enhancement mechanism of small (carbonaceous) grains might be necessary.
For example, rotational (centrifugal) disruption of dust grains by radiative torques would act as an additional
mechanism of small-grain production \citep{Hoang:2019aa}.
However, we note that our models are broadly consistent with the PAH abundances
derived for nearby galaxies (Fig.~\ref{fig:abundance}).
Detailed comparison with the Milky Way MIR emission needs further detailed and extended
modelling, which is left for a future work.

\subsection{Implication for starburst galaxies}\label{subsec:starburst}

\begin{figure}
\begin{center}
\includegraphics[width=0.45\textwidth]{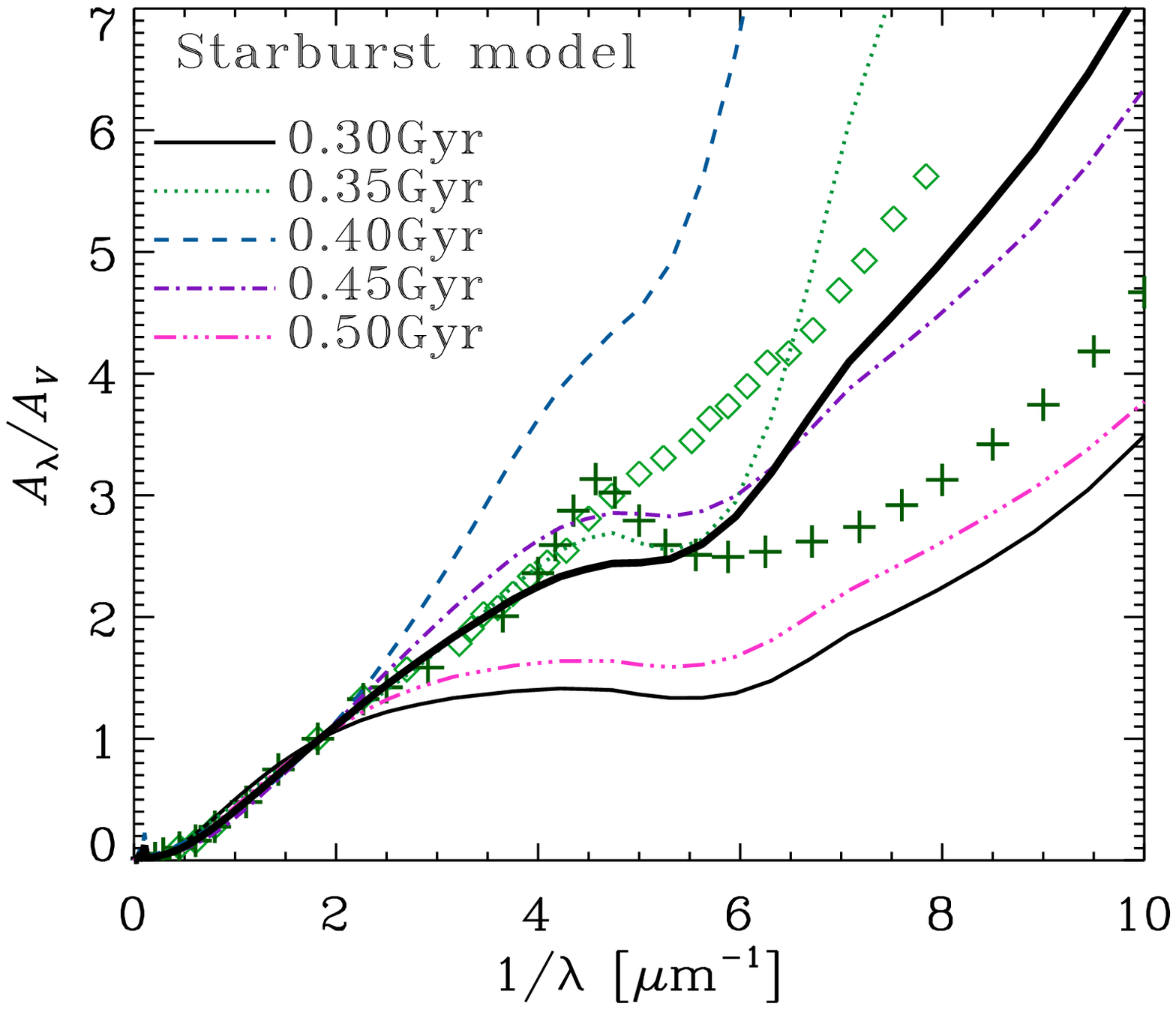}
\includegraphics[width=0.45\textwidth]{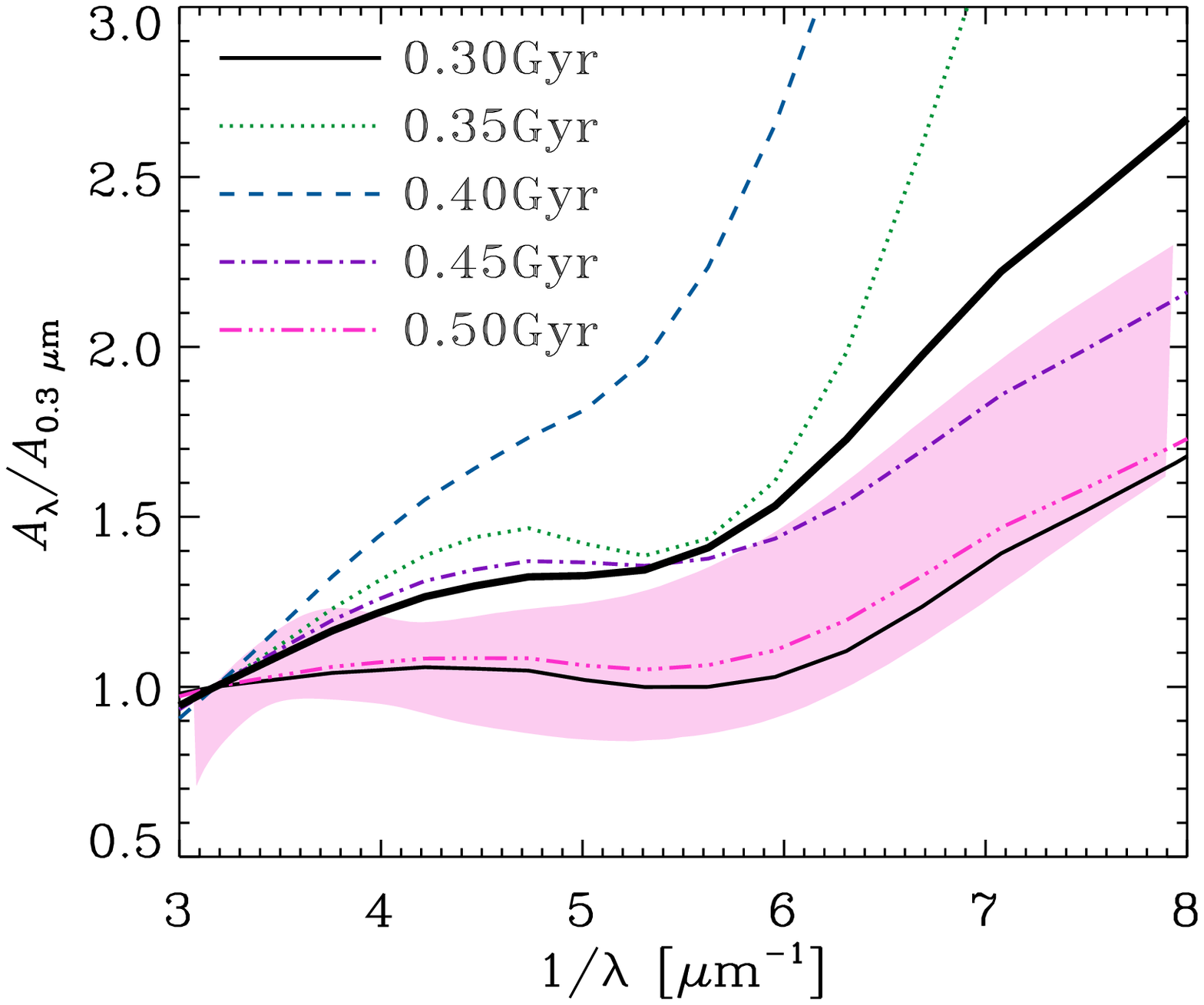}
\end{center}
\caption{Upper: Same as Fig.\ \ref{fig:ext} but for the `starburst model'
(see text) with $\tau_\mathrm{SF}=0.5$~Gyr and $\eta_\mathrm{dense}=0.9$.
We show the evolution of extinction curve in the epoch when the
extinction curve shape drastically changes (at $t=0.3$--0.5 Gyr).
The thin solid, dotted, dashed, dot--dashed, and triple-dot--dashed lines
present the results at $t=0.3$, 0.35, 0.4, 0.45, and 0.5 Gyr, respectively.
The thick solid line shows the averaged extinction curve for these five ages.
Lower: Same extinction curves normalized to the extinction at 0.3 $\micron$
($A_{0.3~\micron}$). The shaded region shows the extinction curve derived
for a $z=6.2$ quasar by \citet{Maiolino:2004aa} with the range showing the
uncertainty.
\label{fig:ext_sb}}
\end{figure}

In the above, the fiducial models are aimed at explaining the moderately
star-forming galaxies in the nearby Universe. On the other hand, we have also
examined the dependence on $\eta_\mathrm{dense}$ and $\tau_\mathrm{SF}$
to extend the predictions to various types of galaxies. Among various types,
starburst galaxies are particularly important from an observational point of view
for the following reasons. First, they are bright in infrared dust emission; thus,
revealing their dust properties is fundamentally important to understand the emission
properties of starburst galaxies. Second, their contribution to the total star formation
activities in the Universe becomes high toward high redshift $z\sim 2$
\citep[e.g.][]{Takeuchi:2005ab,Goto:2010aa,Burgarella:2013aa}.

Because of the above important aspects of starburst galaxies, it is useful
to predict their dust properties using our models. Starbursts could be
characterized by fast star formation in a dense environment.
In our framework, thus, starburst activities could be mimicked by
adopting short $\tau_\mathrm{SF}$ and high $\eta_\mathrm{dense}$.
For comparison with the above results, we adopt the
shortest $\tau_\mathrm{SF}$ (= 0.5 Gyr) and the highest $\eta_\mathrm{dense}$ (=0.9)
to investigate the dust evolution in a starburst. This is referred to as the starburst model.

In Fig.\ \ref{fig:ext_sb}, we show the evolution of extinction curve for the starburst model.
We focus on the
epoch when the extinction curve shape changes drastically by
interstellar processing (especially, accretion and coagulation).
This occurs at $t\sim 0.3$--0.5~Gyr
(i.e.\ time-scales comparable to $\tau_\mathrm{SF}$). At $t=0.3$ Gyr, the extinction curve starts to
become steep because of the small-grain production by shattering and accretion.
The extinction curve is the steepest at $t\sim 0.4$ Gyr and is flatter afterwards
because of coagulation. Note that the variation of extinction curve shown
in Fig.\ \ref{fig:ext_sb} `sandwiches' the SMC extinction curve.
This implies that the SMC extinction curve could be explained by the dust evolution in
starbursts. In addition, the extinction curves are basically bumpless, matching the
characteristics of the SMC extinction curve shape. The low aromatic fraction as well as
the high $f_\mathrm{sil}$ is the cause of the lack of 2175 \AA\ bump.
Since star formation and dust enrichment could occur in a spatially inhomogeneous way in
reality, the extinction curve could be averaged. Thus, for the purpose of presentation,
we show the average of the extinction curves for the five ages shown in Fig.\ \ref{fig:ext_sb}
by the thick line. We observe that the thick line is near to the SMC curve. In particular,
the steep rise without a bump is reproduced. Essentially, the important features of our
starburst model in reproducing the SMC extinction curve are the following two:
(i) rapid modification of grain
size distribution, which occurs on timescales $\sim\tau_\mathrm{SF}$ ($\sim$ metal/dust
enrichment time-scale), is important in explaining the steepness of the SMC extinction
curve; and (ii) the short $\tau_\mathrm{SF}$ and high $\eta_\mathrm{dense}$ keep the
fraction of carbonaceous/aromatic dust low.

We also compare the resulting extinction curves with the one observed
in a quasar at $z=6.2$ \citep{Maiolino:2004aa} as a representative
extinction curve in the epoch when the cosmic age is comparable
to 0.5 Gyr (bottom panel of Fig.\ \ref{fig:ext_sb}). Other quasars at
high redshift ($z\gtrsim 4$) in \citet{Gallerani:2010aa}'s sample show extinction
curves roughly in the shaded region in Fig.\ \ref{fig:ext_sb} (note that the shaded
region shows the uncertainty in the extinction curve of the above quasar).
The observed extinction curve is reproduced well with our model either by
a young age when the dust production is dominated by stellar sources or
by an old age after coagulation has flattened the extinction curve.
\citet{Nozawa:2015aa} explained the same extinction curve with
the later coagulation-dominated phase, while \citet{Maiolino:2004aa},
\citet{Hirashita:2005ab}, and \citet{Bianchi:2007aa} reproduced it with dust produced by SNe.
We note that all these studies assumed all carbonaceous materials to be amorphous
while we actually calculated the fraction of amorphous carbon.
A high dense-gas fraction and
a short star formation time-scale suppress the aromatic fraction, so that
we successfully reproduced extinction curves with no 2175 \AA\ bump.

{}From the above results, we conclude that starburst environments
favour bumpless extinction curves. This is consistent with the bumpless
extinction cuves observed in some actively star-forming environments such as
the SMC and high-redshift quasars.

It is worth mentioning that \citet{Bekki:2015aa} explained the bumpless shapes of
extinction curves by selective loss (outflow) of small carbonaceous grains.
However, they did not actually solve the size-dependent grain motion, so that it
is not clear whether small carbonaceous grains are really selectively lost.
The results provided by this paper give an alternative scenario for the formation
of bumpless extinction curves, and indicate that outflow is not necessarily
needed to explain the bumpless extinction curves.

Finally, we also need to mention that extinction curves may not be easy to obtain
observationally.
For some galaxies, attenuation curves, which include all the radiation transfer
effects, are obtained instead of extinction curves \citep[e.g.][]{Calzetti:2001aa}.
To obtain theoretical attenuation curves, we need to perform radiation
transfer calculations with detailed spatial distributions of dust and stars.
The effects of dust distribution
geometry and stellar-age-dependent extinction make the attenuation curve
significantly different from the original extinction curve
\citep[e.g.][]{Inoue:2005aa,Narayanan:2018aa}.
We leave radiation transfer calculations and comparisons with
observed attenuation curves for future work.

\subsection{Implication for the redshift evolution of PAH abundance}
\label{subsec:z_PAH}

In the above, we have proposed that the carbonaceous dust is dominated
by the non-aromatic component in the starburst environment.
On the other hand, the increase of PAH-to-gas ratio is seen
at a higher metallicity
for a shorter $\tau_\mathrm{SF}$ (Section \ref{subsec:sfh}).
It has been shown that the cosmic star formation activities are dominated
by starburst galaxies at $z\sim 1$--2 \citep[e.g.][]{Takeuchi:2005ab,Goto:2010aa,Burgarella:2013aa}.
Combining these two results, we argue that the PAH abundance is strongly
suppressed at low metallicity ($Z\lesssim 0.1$ Z$_{\sun}$ as seen in our short-$\tau_\mathrm{SF}$
models; Fig.\ \ref{fig:abundance_tau}) in high-redshift galaxies.
Moreover,
the PAH-to-gas ratio is suppressed to a level of $\sim 10^{-5}$--$10^{-6}$
if the gas is dominated by the dense phase
(Fig.\ \ref{fig:abundance}). Therefore,
we should take the suppression of PAH emission into account when we observationally
target PAH emission at high redshift.
The predicted PAH abundances in this paper should be
tested in future observations.
Calculating the SEDs of PAHs and dust based on the
results in this paper would be useful to directly relate the theoretical
predictions to observations.

\subsection{Prospects for hydrodynamic simulations}

Our one-zone model has a limitation in predicting the evolution of the
multi-phase ISM. On the other hand, we have shown that the ISM
phases are important for PAH and dust evolution. Thus, for future work,
it is desirable to predict the evolution of the ISM. Hydrodynamic
simulations provide viable platforms on which dust evolution is calculated
in a consistent manner with the evolution of the ISM
\citep[e.g.][]{McKinnon:2016aa,Zhukovska:2016aa,Hu:2019aa}; some
included the information on grain size distribution
\citep{Aoyama:2017aa,Hou:2017aa,Gjergo:2018aa,Aoyama:2018aa,Hou:2019aa}.
\citet{McKinnon:2018aa} and \citet{Aoyama:2020aa} have already implemented the
evolution of grain size distribution
in hydrodynamic simulations of galaxies.
Semi-analytic approaches also provide an alternative way of modelling the dust formation
in the cosmic structure formation
\citep{Popping:2017aa,Vijayan:2019aa}.
The treatment of multiple
grain species developed in this paper is to be included in hydrodynamic simulations
(and semi-analytic models) for
the purpose of predicting the evolution of PAH abundance and extinction curve.

Spatially resolved information obtained through hydrodynamic simulations is
particularly useful for comparison with observation data.
There are some observations for nearby galaxies, especially for the
Large Magellanic Cloud (LMC) and the SMC, which spatially resolved the
PAH abundance within individual galaxies \citep[e.g.][]{Sandstrom:2012aa}.
In particular, \citet{Chastenet:2019aa} showed that the the PAH abundance in the
diffuse ISM of the LMC is as high as that in the Milky Way in spite of its lower
metallicity. This suggests that the PAH abundance is indeed enhanced in the
diffuse ISM as indicated by our model.
It is also interesting to point out that the abundance of small grains
contributing to the excess emission at 70 $\micron$ is also enhanced
in the diffuse ISM of the LMC \citep{Bernard:2008aa}.
However, \citet{Chastenet:2019aa} also indicated similar PAH
abundances in the dense and diffuse ISM, which imply that the
mixing between these two ISM phases occurs on a short time-scale.
In our framework, the aromatization and aliphatization time-scales
are typically shorter than $10^6$ yr, which means that a mixing time-scale
shorter than $10^6$ yr is necessary to explain the equal PAH abundances between
the diffuse and dense ISM. Such a short mixing time-scale has never been reported.
We suspect that there is still an effect of finite spatial resolution in the observations, which makes it
difficult to fully separate the dense and diffuse ISM. Moreover, there are
dense (molecular) regions which cannot be traced by CO in low-metallicity
galaxies \citep[e.g.][]{Madden:1997aa}, while \citet{Chastenet:2019aa} traced
dense regions by CO emission.
For further comparison, it is interesting to predict the spatially resolved
PAH abundances together with CO emission (an example of predicting the CO
abundance in an entire galaxy is seen in e.g.\ \citealt{Chen:2018aa}) in a hydrodynamic simulation.

\section{Conclusion}\label{sec:conclusion}

We formulate and calculate the evolution of dust and PAHs in a galaxy based on our
evolution model of grain size distribution. We newly separate the grain species
into silicate and carbonaceous dust, and further divide the carbonaceous dust into
aromatic and non-aromatic species. 
To estimate the fractions of various dust species, we calculate the
abundance ratios of silicon to carbon based on the chemical evolution model
and include aromatization and aliphatization (inverse reaction of aromatization).
We regard small aromatic grains in a radius range of 3--50~\AA\ as PAHs.
Since aromatization and aliphatization occur predominantly in
the dense and diffuse ISM, respectively, we introduce the dense gas fraction,
$\eta_\mathrm{dense}$ as a constant parameter.
This fraction also regulates the efficiency of various dust processing mechanisms that
act only in the dense (accretion and coagulation) or diffuse (shattering) ISM.
The star formation time-scale $\tau_\mathrm{SF}$
is also an important parameter since it determines the speed of chemical enrichment in the system. 
We calculate the extinction curves by assuming that the organized carbon structures in aromatic
grains manifest graphite optical properties while the irregular structures in non-aromatic grains
show amorphous carbon properties.

We find that since the time-scales of aromatization and aliphatization are much shorter than the
mass exchange time-scale between the dense and diffuse ISM, the aromatic fraction is simply determined
by the fraction of the diffuse ISM ($1-\eta_\mathrm{dense}$) in most of the grain size range.
This means that the PAH abundance is higher for lower $\eta_\mathrm{dense}$.
In addition, if the ISM is dominated by the diffuse phase, the grain size distribution is biased to
small radii because of efficient shattering and inefficient coagulation. Therefore, the PAH abundance
is sensitive to $\eta_\mathrm{dense}$.

The star formation time-scale ($\tau_\mathrm{SF}$) also affects
the evolution of the PAH abundance.
For shorter $\tau_\mathrm{SF}$, PAH abundance rises only at higher metallicity:
the metallicity level at which a rapid rise of PAH abundance occurs is roughly estimated as
$0.1(\tau_\mathrm{SF}/5~\mathrm{Gyr})^{1/2}$~Z$_{\sun}$.
This metallicity is determined by the value at which dust growth by accretion rapidly
raises the dust abundance.

The extinction curve evolves in the following way for $\tau_\mathrm{SF}=5$~Gyr,
roughly appropriate for nearby spiral galaxies: in the early epoch
($t\lesssim 0.3$ Gyr) when the dust abundance is dominated by
stellar sources, the extinction curve is flat. After that, the dust abundance
is rapidly increased by accretion, which produces a very steep extinction curve,
even steeper than the SMC extinction curve. At $t\gtrsim 3$~Gyr, coagulation makes the
extinction curve flatter, reproducing a Milky-Way-like extinction curve.
The evolution of extinction curve is sensitive to $\eta_\mathrm{dense}$, because, as mentioned
above, the small-grain abundance depends strongly on it.
For small $\eta_\mathrm{dense}(\sim 0.1)$, the extinction curve stays steep at the later
stage, while for large $\eta_\mathrm{dense}(\sim 0.9)$, the extinction curve shows no prominent
bump because of low aromatic fractions.
The extinction curves are also affected by the star formation time-scale ($\tau_\mathrm{SF}$).
A similar shape of extinction curve is realized at
the same value of $t/\tau_\mathrm{SF}^{1/2}$.

Finally, we discuss the implications of our results for starburst galaxies.
We examine the starburst
environment by adopting a `starburst model' in which we adopt a short
$\tau_\mathrm{SF}=0.5$~Gyr and a high $\eta_\mathrm{dense}=0.9$.
In the starburst model,
extinction curves quickly evolve on a time-scale of 0.3--0.5 Gyr, maintaining bumpless shapes.
The range of extinction curves predicted by the starburst model
covers the observed extinction curves of the SMC and high-redshift quasars.
Thus, our model is successful in explaining the variety in extinction curve shapes at
low and high redshifts.

\section*{Acknowledgements}
 
We are grateful to Y.-H. Huang and the anonymous referee for useful comments.
HH thanks the Ministry of Science and Technology for support through grant
MOST 107-2923-M-001-003-MY3 and MOST 108-2112-M-001-007-MY3,
and MSM acknowledges the support from the RFBR grant 18-52-52006.



\bibliographystyle{mnras}
\bibliography{/Users/hirashita/bibdata/hirashita}


\appendix


\bsp	
\label{lastpage}
\end{document}